\documentclass[journal]{IEEEtran}  

\IEEEoverridecommandlockouts      

\usepackage{amsmath}
\allowdisplaybreaks[4]

\renewcommand{\baselinestretch}{1}
\usepackage{graphics} 
\usepackage{epsfig} 
\usepackage{epic,eepic,epsfig}
\usepackage{epstopdf}
\usepackage{mathptmx} 
\usepackage{times} 
\usepackage{amsmath} 
\usepackage{amssymb}  
\usepackage{accents}
\usepackage{bm}
\usepackage{cancel}
\usepackage{mathrsfs}

\usepackage{float}
\usepackage{pstricks}
\usepackage[normalem]{ulem}
\usepackage{hyperref}
\usepackage[capitalize, nameinlink]{cleveref}
\usepackage{algorithm}
\usepackage{algpseudocode}
\usepackage{multirow}
\usepackage{lipsum}
\usepackage{graphicx}
\usepackage{caption}
\usepackage{subcaption}
\usepackage{fancyhdr}   

\def\lungo #1{\mathord{\buildrel{\lower3pt\hbox{$\scriptscriptstyle\frown$}}
\over #1 } }

\def\1D#1#2{{{\partial}\over{\partial #2}}#1}

\def\1d#1#2{{{d}\over{d #2}}#1}

\newcommand{{\R}}{{\mathbb{R}}}
\newcommand{{\C}}{{\mathbb{C}}}

\newtheorem{remark}{Remark}

\begin{document}

\title{A Real-Time Robust Ecological-Adaptive Cruise Control Strategy for Battery Electric Vehicles}

\author{Sheng Yu, Xiao Pan, Anastasis Georgiou, Boli Chen, Imad M. Jaimoukha and Simos A. Evangelou
\thanks{S. Yu, X. Pan, I. M. Jaimoukha and S. A. Evangelou are with the Department of Electrical and Electronic Engineering at Imperial College London, UK
        {\tt\small (sheng.yu17@imperial.ac.uk, xiao.pan17@imperial.ac.uk, i.jaimouka@imperial.\\ac.uk, s.evangelou@imperial.ac.uk)}}%
\thanks{A. Georgiou is with the College of Science \& Engineering at University of Minnesota Twin Cities, USA
        {\tt\small (georg611@umn.edu)}}%
\thanks{B. Chen is with the Department of Electronic and Electrical Engineering at University College London, UK
        {\tt\small (boli.chen@ucl.ac.uk)}}%
}


\thispagestyle{empty}
\setcounter{page}{0}
\begin{figure*}
\centering
\includegraphics[width=\textwidth]{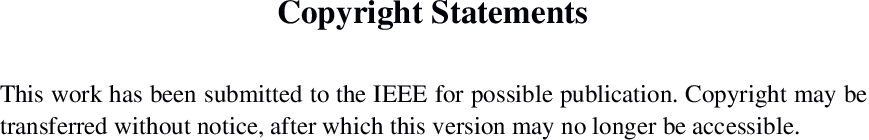}
\end{figure*}

\maketitle
\setlength{\headheight}{22.41992pt}
\thispagestyle{fancy}
\chead{This work has been submitted to the IEEE for possible publication. Copyright may be transferred without notice, after which this version may no longer be accessible.}
\rhead{~\thepage~}
\renewcommand{\headrulewidth}{0pt}

\pagestyle{fancy}
\chead{This work has been submitted to the IEEE for possible publication. Copyright may be transferred without notice, after which this version may no longer be accessible.
}
\rhead{~\thepage~}
\renewcommand{\headrulewidth}{0pt}

\begin{abstract}
This work addresses the ecological-adaptive cruise control problem for connected electric vehicles by a computationally efficient robust control strategy. The problem is formulated in the space-domain with a realistic description of the nonlinear electric powertrain model and motion dynamics to yield a convex optimal control problem (OCP). The OCP is approached by a novel robust model predictive control (RMPC) method handling various disturbances due to modelling mismatch and inaccurate leading vehicle information. The RMPC problem is solved by semi-definite programming relaxation and single linear matrix inequality (sLMI) techniques for further enhanced computational efficiency. The performance of the proposed real-time robust ecological-adaptive cruise control (REACC) method is evaluated using an experimentally collected driving cycle. Its robustness is verified by comparison with a nominal MPC which is shown to result in speed-limit constraint violations. The energy economy of the proposed method outperforms a state-of-the-art time-domain RMPC scheme, as a more precisely fitted convex powertrain model can be integrated into the space-domain scheme. The additional comparison with a traditional constant distance following strategy (CDFS) further verifies the effectiveness of the proposed REACC. Finally, it is verified that the REACC can be potentially implemented in real-time owing to the sLMI and resulting convex algorithm.

\end{abstract}


\section*{Acronyms}
\noindent
\begin{tabular*}{0.49\textwidth}{@{}l@{\hspace{2mm}} @{\extracolsep{\fill}}l}
 CAV & Connected and Automated Vehicle\\
 CDFS & Constant distance following strategy\\
 (R)EACC & (Robust) Ecological Adaptive Cruise Control\\
 (s/m)LMI & (Single/Multiple) Linear Matrix Inequalities\\
 (R)MPC & (Robust) Model Predictive Control\\
 OCP & Optimal Control Problem\\
 RMS & Root-Mean-Square\\
 RSU & Road Side Unit\\
 SDPR & Semi-Definite Programming Relaxation\\
 V2I & Vehicle-to-Infrastructure\\
 V2V & Vehicle-to-Vehicle \\
\end{tabular*}

\section{INTRODUCTION}
With the growing interest in decarbonisation technologies for mitigating urbanisation and environmental issues, intelligent transportation systems with advanced digitalised, automated and electrified road vehicles have been extensively studied \cite{Yuan2022}.
In particular, with the increasing information and intelligence of the transportation field, connected and autonomous vehicles (CAVs) are rapidly developing for the benefits of reduced pollution, increased traffic efficiency, as well as improved driving safety and comfort~\cite{Zhang2019, He2022}. The vehicular ad-hoc networks technology enables CAVs to acquire information of route and other road users through Vehicle to Vehicle (V2V) and Vehicle to Infrastructure (V2I) wireless communications~\cite{Guanetti2018}.
The works in \cite{Saifuzzaman2014, Zhang2023} address the ecological-adaptive cruise control (EACC) problem by real-time optimisation of the driving efficiency of a CAV in a car-following scenario, which is a common driving scenario during everyday driving. 
In this circumstance, the driving behaviour of the ego vehicle is highly dependent on the leading vehicle. A velocity change of the leading vehicle may not be responded to properly by a conventional vehicle operated by a human driver \cite{Ge2017}, thereby leading to unnecessary accelerating/braking or even emergency manoeuvres, which result in additional energy usage \cite{Li2017} and reduced traffic efficiency \cite{Sharma2021}. 
To address the above issues, recent research has focused on developing EACC systems in order to improve safety, energy and traffic efficiencies in car-following scenarios \cite{Zhu2023, Jia2019Trans, PanXiao2020IFAC, Lee2022}. 

The foundation of solving an EACC problem in real-time involves a proper modelling framework of the vehicle dynamics and the design of a computationally efficient EACC control strategy \cite{Peng2023}.
The dynamics model considers both longitudinal dynamics and energy consumption models. 
In the literature, longitudinal dynamic equations can vary from simplified linear models, which exclude any resistive forces \cite{Zheng2014} to the more realistic but nonlinear models due to the presence of the nonlinear friction losses \cite{Zhang2017}. To address the nonlinearity issue, \cite{Hu2020} defines a synthetic control law, where the nonlinear terms can be compensated by feedback linearisation.
In the powertrain modelling aspect, a commonly used energy consumption model is the $L^{2}$-norm of the acceleration (control input). Nevertheless, this simplified model cannot accurately predict energy usage due to the ignorance of the powertrain characteristics \cite{Pan2022, Hadjigeorgiou2023}. 
Alternatively, a battery electric powertrain model is usually taken into account by a quadratic model of the driving force and the velocity, which strikes a balance between modelling accuracy and convexity of the problem~\cite{PanXiao2023, Lacombe2022}. 
In this context, \cite{Lacombe2022} utilises a sequential quadratic programming method to efficiently solve the nonlinear optimisation problem by reorganising the problem variables. Moreover, \cite{PanXiao2023TCST} presents a convex scheme for a signal-free autonomous vehicle intersection crossing problem through a coordinate transformation from time- to space-domain and non-conservative relaxation, which can ensure the consistency between original and convexified problems.  

In terms of the control strategies of the ego vehicle in an EACC problem, there have been numerous efforts reported in the literature, such as fuzzy control, sliding mode control, learning-based control, and MPC \cite{Dong2021, Wang2020, Boddupalli2022, Tang2022, Zhao2020, Lin2020, Feng2021}.  
More specifically, in \cite{Dong2021}, fuzzy control is employed and then an adaptive law is proposed to control the autonomous vehicle system, which guarantees both deterministic as well as fuzzy performances of the system.
An integral sliding mode control strategy is presented in \cite{Wang2020} for EACC systems. The method is coupled with a disturbance observer that estimates unknown uncertainties of the vehicular system.
In \cite{Boddupalli2022}, a machine learning-based controller is proposed, which can predict unexpected vehicular behaviours to achieve a resilient control solution.
In addition, \cite{Tang2022} utilises a deep Q-network algorithm to learn the control strategies for car-following and powertrain energy management with the assistance of a vision-based distance detector.
Furthermore, MPC-based methods are also widely studied and applied in the field.  
Ref \cite{Zhao2020} proposes a stochastic MPC approach with robust chance constraints, which is addressed by solving the dual problem of the original problem based on the strong duality theory and the semi-definite programming relaxation (SDPR) technique.  
Ref \cite{Lin2020} proposes a novel RMPC concept for a multi-objective adaptive cruise control system provided that the additive disturbances are predictable. This method ensures input-to-state stability by imposing an additional quadratic constraint for the stage and terminal costs in the MPC framework. 
Moreover, a tube-based MPC is adopted by \cite{Feng2021} to cope with uncertainties from non-autonomous vehicles by confining the state and input vectors within tightened feasible sets with a high probability. A feed-forward controller is integrated and triggered in the event of unusually large disturbances. 
Earlier initial work of the authors, which serves as a precursor of the current work, utilises an RMPC method with SDPR and multiple linear matrix inequality (mLMI) constraints techniques to make progress with addressing the modelling mismatches in the EACC problem~\cite{Yu2022ECC, Yu2023ICM}.

Despite a rich literature and the previous work of the authors on EACC, there is still a lack of improvement in computational speed for real-time implementation, vehicle and powertrain modelling accuracy for global optimality, and robust guarantee against unavoidable disturbances.
Based on some preliminary results presented in~\cite{Yu2022ECC}, this paper further addresses the concerns on model accuracy, control robustness, and computation efficiency of the EACC problem through designing a space-domain modelling framework, and developing a robust and convex MPC scheme.
Specifically, this paper makes the following contributions:
\begin{itemize}
    \item It proposes a novel real-time robust ecological-adaptive cruise control (REACC) strategy for an electric CAV, which unlike previous work in the literature and \cite{Yu2022ECC, Yu2023ICM} a) utilises a precisely fitted electric powertrain model that considers energy conversion and mechanical transmission losses, b) explicitly defines vehicle dynamic modelling mismatches on air-drag coefficients, tyre-rolling resistance coefficients, and road slope angles rather than adopting Gaussian distributed random disturbances as in previous literature and \cite{Yu2022ECC}, and c) takes into account the communication or sensing error of the leading vehicle by the dynamic model of the vehicle time gap.
    The incorporation of all these practical factors greatly enhances the strategy's potential for practical implementation and can lead to further optimised solutions.
    \item By the choice of the space-domain, unlike previous literature and \cite{Yu2023ICM} in which the time-domain is utilised, it becomes possible to frame the resulting control problem into a convex optimal control problem (OCP), which is solved by the use of LMI optimisation.
    \item Compared to other LMI-based MPC formulations, this paper further makes theoretical contributions required for the present application by suggesting a novel RMPC technique for a system subject to direct additive bounded disturbances, instead of recasting disturbances into uncertainties as in \cite{Georgiou2023}, where problem constraints are captured by a single LMI reformulation rather than addressing constraints element by element as in \cite{Yu2022ECC, Yu2023ICM}. 
    The new sLMI-based REACC method does not significantly sacrifice its robust properties compared to the standard elementwise formulation, while it is showing significant improvement with respect to computational complexity. Importantly, the sLMI formulation enables real-time implementation for fast dynamic systems, such as the proposed REACC that at the same time outperforms existing robust EACC benchmarks in terms of energy consumption and driving comfort, as mentioned next.
    \item The energy consumption and driving comfort performances of the REACC method are investigated and compared with a recently proposed time-domain RMPC method employing an acceleration $L^{2}$-norm based energy consumption model \cite{Yu2023ICM}, by comprehensive numerical case studies. The proposed method demonstrates a $4.5\% {\sim} 5\%$ energy saving with also a more comfortable travel experience. Moreover, by comparing with a benchmark strategy using non-optimised cruise control, the constant distance following strategy (CDFS) \cite{Guo2023}, the proposed REACC reduces energy losses by approximately $11\%$.
\end{itemize}

The rest of the paper begins with a statement of the original non-convex EACC problem and modelling in Section~\ref{sec:system description}, followed by Section~\ref{sec:formulations}, which reformulates the original problem into a convex and nominal (disturbance-free) OCP through proper convex relaxation and approximation methods. Moreover, the OCP is rewritten into a condensed nominal MPC format as a benchmark. The proposed RMPC algorithm with SDPR and sLMI (the REACC), which addresses disturbances including the modelling mismatches, is further introduced in Section~\ref{sec:rmpc}. Simulation results of the REACC and benchmark comparisons are illustrated and discussed in Section~\ref{sec:simulation results}. Finally, conclusions are provided and a future work plan is suggested in Section~\ref{sec:conclusions}.

\emph{Notation}: Let $\mathbb{R}$, $\mathbb{R}_{\geq 0}$, $\mathbb{R}_{> 0}$ and $\mathbb{N}_{>0}$ denote the real, the non-negative real, the strict positive real sets of numbers, and non-zero natural numbers, respectively.
$\mathbb{R}^n$ denotes the space of a $n$-dimensional real (column) vector, $\mathbb{R}^{n\times m}$ denotes the space of a $n \times m$ real matrix and $\mathbb{D}^{n}$ denotes the space of a diagonal matrix in $\mathbb{R}^{n \times n}$. 
${I}^{n}$ denotes an $n \times n$ square matrix with ones on the main diagonal and zeros elsewhere. 
$\mathbf{0}^{n\times m}$ denotes an $n \times m$ matrix with all zeros.
$A^{\top}$ represents the transpose of $A$. 
Let a symmetric matrix $Q \in \mathbb{R}^{n\times n}$ with $Q \succeq 0$ denote a positive semi-definite matrix.
For matrices, $Q_1,\dots,Q_n$, $\text{diag}\{Q_1,\dots,Q_n\}$ represents a block diagonal matrix with $Q_{i}$ the $i^{th}$ diagonal matrix.
\section{Statement of EACC Problem }
\label{sec:system description}
This work focuses on the EACC paradigm, where there is an ego CAV (also known as the controlled vehicle) and its driving behaviour is constrained by the traffic in front.
\begin{figure}[t!]
\centering
\includegraphics[width=\columnwidth]{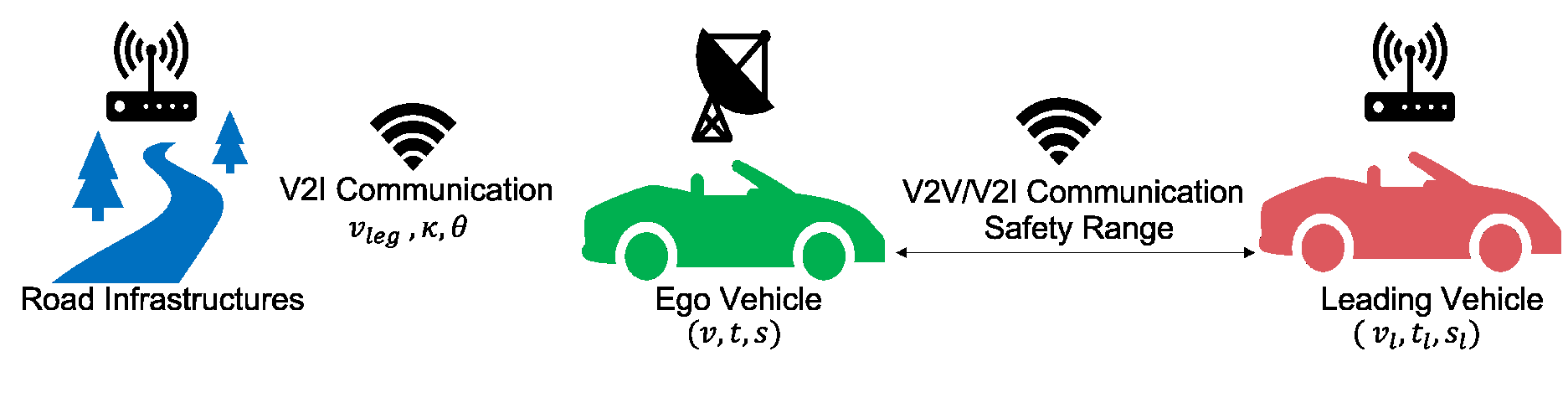}\\[-1ex]
\setlength{\belowcaptionskip}{-15pt}
\caption{Scheme of EACC paradigm with V2I and V2V Communications.}
\label{fig:V2IV_time}
\end{figure}
As it can be seen in Fig.~\ref{fig:V2IV_time}, an ego vehicle is able to acquire real-time road information from GPS or roadside units (RSUs), such as legal speed limit, road curvature, and slope angle, \cite{Jones2019, Xu2022}.
The front traffic can be reasonably formulated as a leading vehicle \cite{Sun2015}, and in the present framework, it is assumed that a speed profile of the leading vehicle is available for the ego vehicle, which can be obtained from the leading vehicle via V2V~\cite{Hyeon2022} or from the RSUs via V2I~\cite{He2021} communication.  
Furthermore, it is assumed that no lane changing or overtaking of the leading vehicle is taking place.

In this paper, the ego vehicle is requested to travel the same distance as the leader, which is predefined. However, the following distance gap is not fixed and can vary within a specified range. In this regard, the aim is to design a real-time robust MPC-based EACC strategy that optimises energy consumption with free-end time. In order to make vehicle travel time as a state variable that can be easily optimised, we formulate the problem in the space-domain \cite{PanXiao2023} rather than in the time-domain as with the majority of optimisations in adaptive cruise control in the literature.
Later on, it will also be shown that the space-domain modelling approach can yield a convex program without sacrificing optimality in terms of the energy economy. 

Let us first denote $s$ the vehicle travelled distance, which is the independent variable in the space-domain formulation. Then, the motion of the ego vehicle can be described by the following dynamic equation \cite{PanXiao2023} 
\begin{equation}
    \frac{d}{ds}E(s)=F_w(s)-F_d(s)-F_r(s)-F_g(s),\label{eq:space_E_continuous}
\end{equation}
where $E(s) = \frac{1}{2}mv(s)^2$ is the kinetic energy of the ego vehicle, with $v(s)$ the velocity of the ego vehicle and $m$ the ego vehicle mass, and $F_w(s)$ is the total force acting on the wheels. Moreover, $F_d(s)=\frac{2f_d(s)E(s)}{m}$ is the air-drag resistance, $F_r(s)=mgf_r(s)\cos(\theta(s))$ is the tyre-rolling resistance, with $g$ the acceleration of gravity and $f_d(s)$ and ${f}_r(s)$ the space-dependent coefficients of air-drag and tyre-rolling resistance forces, respectively. Finally, $F_g(s)=mg\sin(\theta(s))$ is the gradient force due to the road slope angle $\theta(s)$. Without loss of generality, it is assumed that the nominal values of $f_d(s)$, ${f}_r(s)$ and $\theta(s)$ are known from the vehicle characteristics, GPS, and so on. Therefore, the real values of $f_d(s)$, ${f}_r(s)$ and $\theta(s)$ can be represented as 
\begin{equation} \label{eq:space_define_modelling_mismatches}
    \begin{aligned}
       & {f_d}(s)=\tilde{f_d}+\Delta f_d(s),\\
       & {f_r}(s)=\tilde{f_r}+\Delta f_r(s),\\
       & {\theta}(s)=\tilde{\theta}(s)+\Delta \theta(s),
   \end{aligned}
\end{equation}
where $\tilde{f_d}$, $\tilde{f_r}$, and $\tilde{\theta}$ are the nominal parameters available to CAVs, and $\Delta f_d(s)$, $\Delta f_r(s)$, and $\Delta \theta(s)$ are the unknown parts, treated as modelling mismatch. 
The mismatch of the real to the nominal air-drag and tyre-rolling resistance coefficients, respectively, is assumed to be bounded with ${f}_d(s) \in [\underline{{f}}_{d},\overline{{f}}_{d}]$ and ${f}_r(s) \in [\underline{{f}}_{r},\overline{{f}}_{r}]$.
The bounds $\underline{{f}}_{d}$ and $\overline{{f}}_{d}$ of the air-drag resistance coefficient can be determined with reference to the physical relationship between the air-drag resistance coefficient and the headway distance~\cite{PanXiao2020IFAC,Lopes2019}. The tyre-rolling resistance coefficient limits, $\underline{{f}}_{r}$ and $\overline{{f}}_{r}$, are determined based on the investigation of practical tyre-rolling coefficients at the International Organization for Standardization (ISO) conditions~\cite{Ejsmont2017}. 
Furthermore, $\Delta \theta(s)$ represents the gap between the real road slope angle $\theta(s)$ and the nominal angle data $\tilde{\theta}(s)$ collected by road infrastructures including RSUs (which are accessible to CAVs) at the position $s$. This modelling mismatch on road gradients can be caused by RSU measuring errors, speed humps or other temporary road work.
In the present work, the gradient mismatch range is assumed to be a bounded disturbance with $\Delta \theta(s)\in [\underline{\Delta \theta},\overline{\Delta \theta}]$.

Next, considering $v_l(s)$ the leading vehicle velocity, the dynamics of the time headway between the two vehicles, $\Delta t(s)$, are governed by 
\begin{equation}\label{eq:space_t_continuous}
    \frac{d}{ds}\Delta t(s)=\frac{1}{v(s)}-\frac{1}{v_l(s)}.    
\end{equation}
For the sake of further discussion and the introduction of the RMPC framework, the system \eqref{eq:space_E_continuous}-\eqref{eq:space_t_continuous} is discretised by forward Euler discretisation subject to a sampling interval $\delta s\in\mathbb{R}_{>0}$, leading to the discrete dynamic system 
\begin{subequations}
\begin{align}
 &  E(k+1)= E(k) +\bigg(F_w(k)-\frac{2{f}_d(k)E(k)}{m} \nonumber\\
  & \hspace{1cm} -mg{f}_r(k)\cos\left({\theta}(k)\right)\!-\!mg\sin\left({\theta}(k)\right)\bigg)\delta s,
  \label{eq:space_E_accurate}\\
  &   \Delta t(k+1)=\Delta t(k)+\left(\frac{1}{\sqrt{2E(k)/m}}-\frac{1}{{v_l}(k)}\right)\delta s, \label{eq:space_dt_accurate}
    \end{align}
\end{subequations}
where the sampling index $k\in \mathbb{N}_{[0, k_s]}$ with the total number of samples $k_s=S_f/\delta s\in\mathbb{N}_{>0}$ ($S_f$ is the predefined total travelled distance). The boundaries of the permissible range of the time gap are constructed below
\begin{equation}
    \begin{aligned}
    \Delta t_{\min}\leq \Delta t(k) \leq \Delta t_{\max},\label{eq:space_dt_constraint}
    \end{aligned}
\end{equation}
where $\Delta t_{\min}$ is the minimum time gap to avoid rear-end collision, and $\Delta t_{\max}$ is the maximum allowed time gap to improve traffic capacity and maintain adequate V2V/V2I communication.
For safety purposes, the kinetic energy $E(k)$ is bounded by
\begin{equation}
    \begin{aligned}
    E_{\min}\leq E(k)\leq  E_{\max}(k), \label{eq:space_dE_constraint}
    \end{aligned}
\end{equation}
where $E_{\min}= {\frac{1}{2}mv_{\min}^2}$ and $E_{\max}(k)=\frac{1}{2}mv_{\max}^2(k)$ are the lower and upper bounds of kinetic energy, determined by the minimum allowed speed
$v_{\min}$, which is a sufficiently small positive constant aiming to avoid the singularity issues in \eqref{eq:space_t_continuous} without sacrificing the generality of the formulation, and the maximum speed limit $v_{\max}(\kappa(k))$. Note that $v_{\max}(\kappa(k))$ is modelled as a function of the real-time road curvature $\kappa(k)$ to ensure safety and comfort during cornering. It is estimated by the concept of the acceleration diamond \cite{ChenBoli2019} that represents a combined longitudinal and lateral acceleration constraint for ordinary driving behaviour
\begin{equation}
    \begin{aligned}
    \bigg |\frac{F_{w}(k)/m}{a_{x,\max}}\bigg|+\bigg|\frac{v(k)\Omega(k)}{a_{y,\max}}\bigg|\leq 1, \label{eq:acceleration diamond}
    \end{aligned}
\end{equation}
where $\Omega(k)={v(k)}{\kappa(k)}$ represents the yaw rate of the vehicle, and $F_{w}(k)/m$ and $v(k)\Omega(k)$ are the longitudinal and lateral accelerations, and their individual limits are denoted by $a_{x,\max}$ and $a_{y,\max}$, respectively. By reorganising \eqref{eq:acceleration diamond}, the maximum cornering velocity limit can be calculated by
\begin{equation}
    v_{\text{corner},\max}(k)=\sqrt{\bigg(1-\frac{F_{w,\max}/m}{a_{x,\max}}\bigg)\frac{a_{y,\max}}{\kappa(k)}},
    \label{eq:v_limit_cornering}
\end{equation}
in which $F_{w,\max}$ is the maximum force that can be provided by the vehicle powertrain (see \eqref{eq:nonlinear_Fw_constraint_discrete} below) to the wheels.
Thus, the constraints \eqref{eq:space_dE_constraint} on $E(k)$ can be specified as follows 
\begin{equation}
    v_{\min} \leq v(k) \leq \underbrace{\text{min}\bigg(v_{leg}(k),\sqrt{\bigg(1-\frac{F_{w,\max}/m}{a_{x,\max}}\bigg)\frac{a_{y,\max}}{\kappa(k)}} \bigg)}_{v_{\max}(k)}, \label{eq:nonlinear_v_constraint} 
\end{equation}
where the upper-speed limit, $v_{\max}(k)$, merges the legal speed limit of the road, $v_{leg}(k)$, and the cornering speed limit \eqref{eq:v_limit_cornering}. During the driving task, the ego CAV can be informed $v_{\min}$ and $v_{\max}(k)$ ($v_{leg}(k)$ and $\kappa(k)$) from the infrastructure through the V2I communication. 

\begin{figure}[t!]
\centering
\includegraphics[width=\columnwidth]{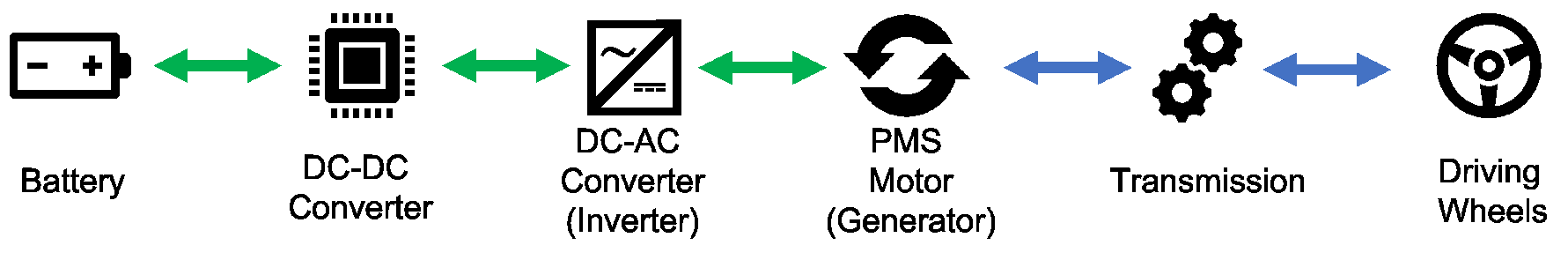}\\[-1ex]
\caption{Block diagram of the battery electric vehicle powertrain with a DC-DC converter, a DC-AC converter (an inverter), a PMS motor (a generator), and a mechanical transmission set. Green and blue arrows represent electrical and mechanical power flows, respectively.}
\label{fig:powertrain_block}
\end{figure}

Moreover, the ego vehicle is assumed to be equipped with a battery-electric powertrain, which is illustrated in Fig.~\ref{fig:powertrain_block}.
The powertrain connects the battery (energy source) to the driving wheels (loads) through a series of components including a DC-DC converter, a DC-AC converter (an inverter), a permanent magnet synchronous (PMS) machine (motor/generator), and a mechanical transmission set that delivers the powertrain driving force, $F_t$, to the wheels. Both the converters and the transmission set can be simply modelled by constant efficiency factors \cite{ChenBoli2019} (see Table. \ref{tab:model_specification_parameters}), while the efficiency of the PMS machine is modelled as a static efficiency map from ADVISOR \cite{Markel2002}, with the machine characteristics shown in Table. \ref{tab:model_specification_parameters}. Also, an equivalent circuit model with a constant open circuit voltage, while considering the battery internal resistance, is utilised to model the battery (battery specifics are provided in Table. \ref{tab:model_specification_parameters}) \cite{ChenBoli2019}. 
\begin{table}[ht]
    \centering
        \caption{Battery Electric Vehicle Powertrain Main Parameters. 
        }
    \label{tab:model_specification_parameters}
    \begin{tabular*}{1\columnwidth}{c @{\extracolsep{\fill}} c@{\extracolsep{\fill}}c}
        \hline
        \hline
          Description & Symbols & Values \\
         \hline
         Battery open circuit voltage &  & 432 V \\

         Battery internal resistance &  & 0.12 $\Omega$ \\

         Battery package overall capacity &  & 20.736 kWh\\

         DC-DC converter efficiency &  & 0.97  \\
    
         DC-AC converter efficiency &  & 0.96  \\
   
         PMS machine state resistance & & 90 $\Omega$\\

         PMS machine rotor magnetic flux &  & 0.21 Wb\\

         PMS machine number of poles &  & 6 \\

         Transmission efficiency &  & 0.96\\

         Minimum powertrain driving force & $F_{t,\min}$ & -3500 N\\ 
 
         Maximum powertrain driving force & $F_{t,\max}$ & ~3500 N\\

         Largest mechanical braking force & $F_{m,\min}$ & -4300 N\\ 

        \begin{tabular}{@{}c@{}}Fitted coefficients \\ of battery terminal power \end{tabular} & $a_1/a_2/a_3$ & \begin{tabular}{@{}c@{}}$6.31\times10^{-5}/$\\ $1.046/115.2$ \end{tabular}\\
        \hline
        \hline
    \end{tabular*}
\end{table}

The total force applied on the wheels $F_w(k)$ consists of the powertrain driving force $F_{t}(k)\!\in [F_{t,\min},F_{t,\max}]$ and the non-regenerative (dissipative) mechanical braking force $F_{m}(k)\in[F_{m,\min},0]$ such that
\begin{equation}
    F_{w}(k)=F_{t}(k)+F_{m}(k),\label{eq:nonlinear_Fw_defination}
\end{equation}
and $F_w(k)$ is subject to the following constraints
\begin{equation}
F_{t,\min}+F_{m,min}\leq F_{w}(k)\leq F_{t,\max},
\label{eq:nonlinear_Fw_constraint_discrete}
\end{equation}
with $F_{t,\max}$ ($=\!\!F_{w,\max}$) and $F_{t,\min}$ the maximum traction force and largest (negative) regenerative braking force, respectively, delivered by the electric machine at the wheels, and $F_{m,\min}$ the largest (negative) mechanical braking force at the wheels, assuming that the tyres can provide this range of forces. 

\begin{figure}[t!]
\centering
    \begin{subfigure}{\columnwidth}
         \centering
         \includegraphics[width=\columnwidth]{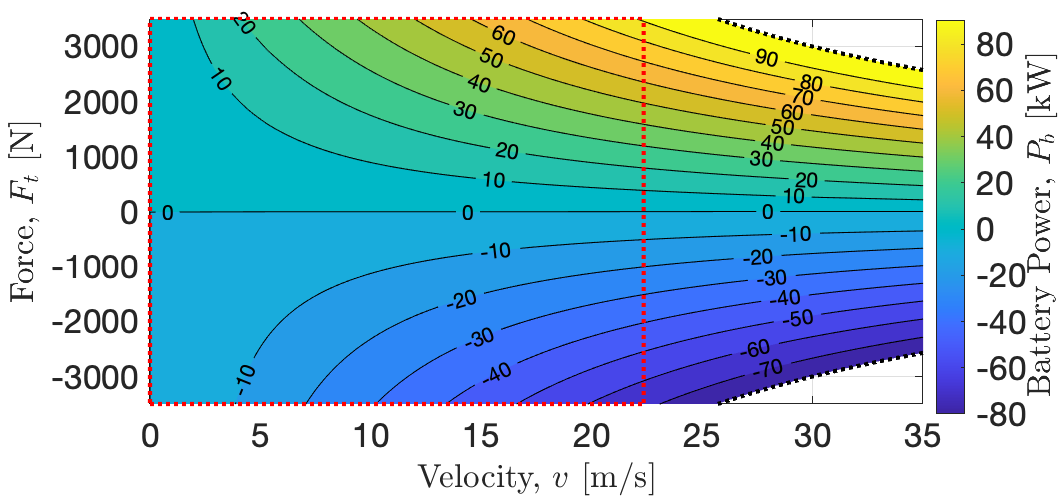}
         \caption{Battery electric vehicle power consumption map.}
         \label{fig:drive_powertrain_map}
    \end{subfigure}
    \begin{subfigure}{\columnwidth}
         \centering
         \includegraphics[width=\columnwidth]{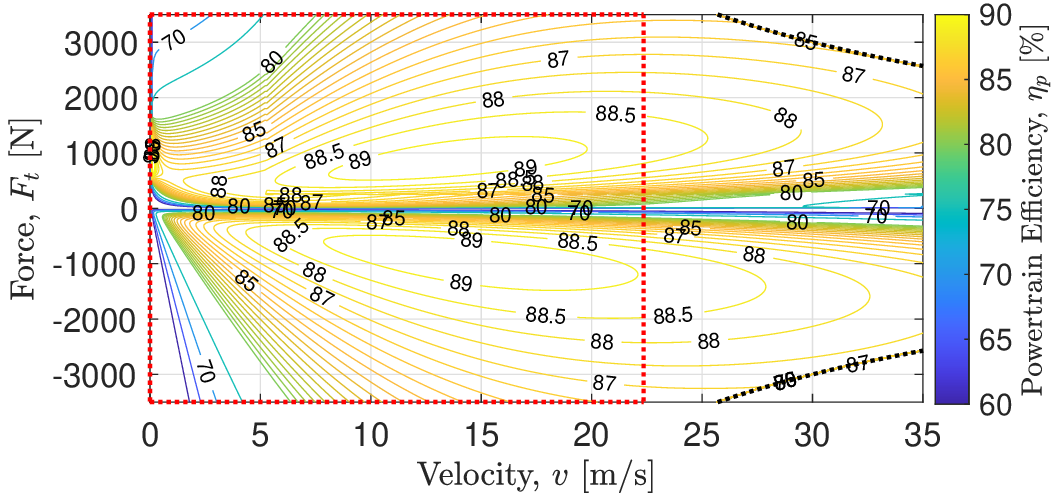}
         \caption{Battery electric vehicle powertrain efficiency map.}
         \label{fig:drive_powertrain_eff}
     \end{subfigure}
\caption{Battery electric vehicle power consumption, $P_{b}$ (shown in Fig. \ref{fig:drive_powertrain_map}) and powertrain efficiency, $\eta_{p}$ (shown in Fig. \ref{fig:drive_powertrain_eff}), associated with powertrain driving force, $F_t$ and vehicle velocity, $v$. Positive force means the battery is discharging while negative force indicates the battery is recharging. Solid lines in Fig. \ref{fig:drive_powertrain_map} and Fig. \ref{fig:drive_powertrain_eff} are battery power consumption and efficiency contour lines, respectively. Black dashed lines are motor operational bounds. Red dashed rectangles denote the feasible overall vehicle powertrain operating range which is determined by the minimum and maximum powertrain driving forces ($F_{t,\min}$ and $F_{t,\max}$), as well as the minimum and maximum velocities ($v_{\min}$ and $v_{\max}$, where for illustration purposes the highest value of $v_{leg}$ is shown for $v_{\max}$ in Fig. \ref{fig:drive_powertrain}). }
\label{fig:drive_powertrain}
\end{figure}

Therefore, the energy consumption of the battery electric vehicle can be evaluated by its battery energy usage, whose rate is the input power drawn from the battery to drive the vehicle, $P_b$. The power $P_b$ can be further represented as a function of $F_t$ and ego vehicle velocity $v$, $P_{b}(F_{t}(k),v(k))$, which is shown in Fig. \ref{fig:drive_powertrain_map}.

The overall electrical-mechanical power conversion efficiency factor $\eta_{p}$ of the battery-electric powertrain is defined as
\begin{equation}
    \begin{aligned}
    \eta_{p}(k)= \left\{
 \begin{array}{ll}
   \displaystyle \frac{F_t(k) v(k)}{P_b(k)}, & \forall F_{t}(k)\geq0, \\
   \\
   \displaystyle
    \frac{P_b(k)}{F_t(k) v(k)}, & \forall F_{t}(k)<0,
 \end{array}\right.        
    \end{aligned}
\end{equation}
which will be employed in Section \ref{sec:simulation results} to evaluate the ecological performance of the proposed method. Fig.~\ref{fig:drive_powertrain_eff} illustrates the overall powertrain efficiency map of $\eta_{p}$, together with the operation limits of the electric machine and powertrain utilised in this work, respectively.

\begin{figure}[t!]
\centering
\includegraphics[width=\columnwidth]{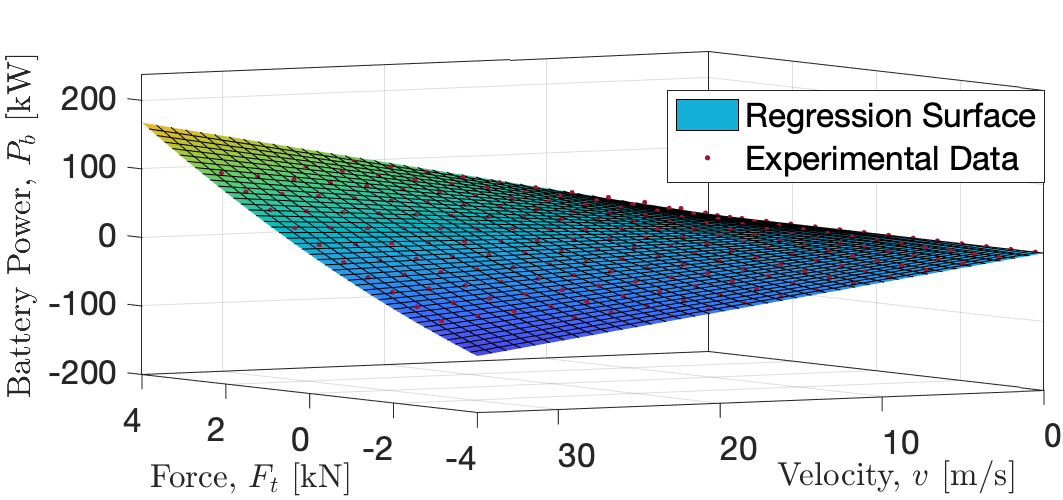}
         \caption{Electric vehicle battery power fitting map by the quadratic function in \eqref{eq:powertrain_fit}. Nonlinear regression of the battery-drawn power data represented by the blue regression surface, calculated based on the power consumption map shown in Fig.~\ref{fig:drive_powertrain_map}, with the coefficient of determination, $R^2= 0.995$.}
         \label{fig:drive_map_fitting}
\end{figure}
Moreover, it is worth noting from Fig.~\ref{fig:drive_powertrain_map} that the battery terminal power $P_b$ can be precisely fitted by a quadratic function of $F_t$ and $v$, as shown in Fig.~\ref{fig:drive_map_fitting}, with $R^2$ of 0.995. 
\begin{equation}
    \begin{aligned}
    P_{b}(k)=\left(a_1F_t(k)^2+a_2F_t(k)+a_3\right)v(k), \label{eq:powertrain_fit}
    \end{aligned}
\end{equation}
where $a_1$, $a_2$ and $a_3$ are the fitted coefficients, which are provided in Table. \ref{tab:model_specification_parameters}. 

The proposed REACC method in this work aims to save travelling time meanwhile reducing energy consumption. Therefore, a tentative multi-objective stage cost function of the EACC problem is designed as follows
\begin{equation}
\begin{aligned}
\label{eq:space_multi_objective0}
    L(k)=& W_{E}\!\! \left(E(k)-E_{\max}(k)\right)^2\!+\! W_{F} \frac{P_{b}(k)}{v(k)},
\end{aligned}
\end{equation}
where $W_{E}, W_{F}\in \mathbb{R}_{>0}$ are two weighting factors. After substituting \eqref{eq:powertrain_fit} into \eqref{eq:space_multi_objective0}, the cost function becomes
\begin{equation}
\begin{aligned}
\label{eq:space_multi_objective}
    L(k)\!=\! W_{E}\!\! \left(E(k)\!-\!E_{\max}(k)\right)^2\!+\!W_{F}\!\left(\!a_1F_t(k)^2\!+\!a_2F_t(k)\!+\!a_3\!\right)\!,
\end{aligned}
\end{equation}
which is a quadratic cost function in terms of $E(k)$ and $F_t(k)$, enabling the formulation of a convex problem in the next section.  
In particular, the first term in the cost function encourages the ego vehicle to follow $v_{\max}$ to maximise mobility whereas the second term aims to minimise the battery energy usage.
\section{Convex Optimal Control Problem Formulation for Nominal EACC}
\label{sec:formulations}
This section formulates the nominal (disturbance-free) EACC problem as an OCP. Owing to the proper convexification techniques, the OCP can be formulated as a convex problem. The convex modelling framework is reorganised into a nominal MPC benchmark and it is then used to design the computationally efficient RMPC in Section~\ref{sec:rmpc}.
\subsection{Convex OCP formulation} 
\label{sec:formulations_OCP}
For the sake of further discussion, let us rewrite the system equation \eqref{eq:space_E_accurate} in the following form with nominal dynamics only
\begin{equation}
\begin{aligned}
     \tilde{E}(k+1)=& \tilde{E}(k) +\bigg((F_w(k)-\frac{2\tilde{f_d}\tilde{E}(k)}{m}\\ 
     &-mg\tilde{f_r}\cos\left(\tilde{\theta}(k)\right)-mg\sin\left(\tilde{\theta}(k)\right)\bigg)\delta s, \label{eq:space_E_tilde}
     \end{aligned}  
\end{equation}
where $\tilde{E}$ is the nominal kinetic energy state. Moreover, the propagating equation of the nominal time headway gap $\Delta \tilde{t}$, extracted from the realistic dynamics in \eqref{eq:space_dt_accurate}, can be rewritten as
\begin{equation}
\label{eq:space_delta_t_tilde}
    \Delta \tilde{t}(k+1)=\Delta \tilde{t}(k)+\left(\frac{1}{\sqrt{2\tilde{E}(k)/m}}-\frac{1}{\tilde{v}_l(k)}\right)\delta s,
\end{equation}
where $\tilde{v}_l(k)$ is the nominal velocity of the leading vehicle available to the ego vehicle, and may not be identical to the actual velocity $v_l(k)$ because of the communication or sensing error.
To deal with the nonlinearity of $\frac{1}{\sqrt{2\tilde{E}(k)/m}}$ existing in the dynamics of \eqref{eq:space_delta_t_tilde} due to the space-domain formulation, an auxiliary variable denoted as $\zeta(k)$ is introduced to convexify the nonlinearity. By defining $\zeta(k)=\frac{1}{\sqrt{2\tilde{E}(k)/m}}$, the dynamics of the time gap \eqref{eq:space_delta_t_tilde} can be relaxed as an equality constraint and a convex path constraint
\begin{subequations}
    \begin{align}
    \Delta \tilde{t}(k+1)&=\Delta \tilde{t}(k)+\left(\zeta(k)-\frac{1}{\tilde{v}_l(k)}\right)\delta s, \label{eq:space_dt_zeta}\\
    \zeta(k)&\geq \frac{1}{\sqrt{2\tilde{E}(k)/m}}. \label{eq:space_zeta_constraint}
    \end{align}
    \label{eq:space_zeta}
\end{subequations}
$\!\!\!$Note that the feasibility of the original nominal state dynamics \eqref{eq:space_delta_t_tilde} is enlarged in \eqref{eq:space_dt_zeta} due to the inequality path constraint \eqref{eq:space_zeta_constraint}. The equivalence between \eqref{eq:space_delta_t_tilde} and \eqref{eq:space_zeta} is valid if the equality of \eqref{eq:space_zeta_constraint} holds for all $k$, which can be ensured under the proposed framework and will be discussed below when introducing the stage cost function. 

By collecting \eqref{eq:space_E_tilde} and \eqref{eq:space_dt_zeta}, a nominal and convex dynamic state-space representation can be summarised as
\begin{equation}
    \begin{aligned}
    \label{eq:space_state_function_set}
    &\tilde{x}(k+1)=A\tilde{x}(k)+B_{u}u(k)+B_{c}C(k),\\
    & A=\begin{bmatrix}
    1-\frac{2\tilde{f_d}}{m}\delta s & 0\\0 & 1
    \end{bmatrix},\,B_{u}= \begin{bmatrix}
    \delta s & 0\\0 & \delta s
    \end{bmatrix},\, \\
    & B_{c}\!=\! \begin{bmatrix}
    \delta s & 0\\0 & \delta s
    \end{bmatrix}\!, \,C(k) \!=\! \begin{bmatrix}
    -mg\tilde{f_r}\cos{\tilde{\theta}(k)-mg\sin{\tilde{\theta}(k)}}\\-\frac{1}{\tilde{v}_l(k)}
    \end{bmatrix}\!,
    \end{aligned}
\end{equation}
where $\tilde{x}(k)\!=\![\tilde{E}(k),\Delta \tilde{t} (k)]^{\top}$ is the nominal state vector, $u(k)\!=\![F_w(k),\zeta(k)]^{\top}$ is the control input. Note that $C(k)$ depends only on $k$ since both the nominal road slope angle $\tilde{\theta}(k)$ and the available leading vehicle speed $\tilde{v}_{l}(k)$ contained by $C(k)$ are varying with $k$.
Furthermore, nominal constraints of the states and the $F_w(k)$ input can be given by
\begin{equation}
    \underline{f}(k) \leq \tilde{f}(\tilde{x}(k),u(k))\leq \overline{f}(k)
\end{equation}
in which $\tilde{f}(k)$ collects \eqref{eq:space_dt_constraint} \eqref{eq:space_dE_constraint}, and  \eqref{eq:nonlinear_Fw_constraint_discrete} as given below
\begin{equation}
    \begin{aligned}
    \label{eq:space_constraint_function_set}
    \tilde{f}(k)&=C_{f} \tilde{x}(k)+D_{fu}u(k),\\
     \tilde{f}(k)&\!=\!\begin{bmatrix}
     \tilde{E}(k)\\\Delta \tilde{t} (k)\\F_w(k)
    \end{bmatrix}\!, C_{f}\!\!=\begin{bmatrix}
    1 & 0 \\ 0 & 1\\0 & 0
    \end{bmatrix}\!, D_{fu}\!=\! \begin{bmatrix}
    0&0\\0&0\\1&0
    \end{bmatrix}, 
    \end{aligned}
\end{equation}
with $\tilde{f}(k)$ bounded by lower and upper constraints $\underline{f}(k)=[ E_{\min}, \Delta t_{\min},F_{t,\min}+F_{m,min}]^{\top}$ and $\overline{f}(k)=[ E_{\max}(k), \Delta t_{\max},\linebreak F_{t,\max}]^{\top}$, respectively. 

Moreover, in order to guarantee the equality condition of \eqref{eq:space_zeta_constraint} once the proposed OCP is formulated, the nominal form of the stage cost \eqref{eq:space_multi_objective} is modified as follows
\begin{equation}
    \begin{aligned}
    \tilde{L}(k)=& W_{E}\left(\tilde{E}(k)-E_{\max}(k)\right)^2 \\
    &+ W_{F}a_1F_w(k)^2 +W_{F}a_2F_w(k) +W_{\zeta}\zeta(k),
    \label{eq:space_multi_objective_convex_stage}
    \end{aligned}
\end{equation}
where an additional cost $W_{\zeta}\zeta(k)$ is introduced, with $W_{\zeta}\in\mathbb{R}_{>0}$ the corresponding weighting factor, the term $a_3$ in \eqref{eq:space_multi_objective} is removed since it is a constant term, and $F_t(k)$ is substituted by $F_w(k)$, which are equivalent if there is no mechanical friction braking; see \eqref{eq:nonlinear_Fw_defination}. The proof can be referred to the authors' previous work in \cite{PanXiao2023TCST}, which involves a similarly convexified stage cost for a different application, and is therefore omitted.
As the powertrain driving force $F_t(k)$ is replaced by $F_w(k)$ that also includes the mechanical friction braking force, the optimality of the convex optimisation problem that will be formulated may be compromised in case the friction braking force is active during the mission (i.e., $\exists k, \,\, F_m(k) \ne 0$). However, friction braking is naturally suppressed in eco-driving to maximise energy recovery, which will also be confirmed by the simulation results in Section \ref{sec:simulation results}.

Based on \eqref{eq:space_state_function_set}--\eqref{eq:space_multi_objective_convex_stage}, the overall nominal EACC problem can be formulated as a convex OCP in the space domain as follows 
\begin{subequations}\label{eq:space_domain_problemformulation}
    \begin{align}
    \min\limits_{u(k)}\quad&\sum_{k=0}^{k_s}\tilde{J}(k)\delta s=\left(\sum_{k=0}^{k_s-1} \tilde{L}(k)\delta s+\tilde{\Psi}(k_s)\right)\, ,
    \label{eq:space_domain_problem_cost_function}\\
    \textbf{s.t.   } &\tilde{x}(k+1)=A\tilde{x}(k)+B_{u}u(k)+B_{c}C(k),
    \label{eq:space_domain_problem_state}\\
    & \underline{f}(k) \leq \tilde{f}(\tilde{x}(k),u(k))\leq \overline{f}(k)\,, \label{eq:space_domain_problem_constraint}\\
    & \zeta(k)\geq \frac{1}{\sqrt{2\tilde{E}(k)/m}}\,, \label{eq:space_domain_problem_constraint_zeta}\\
    \textbf{given:   }&\tilde{x}(0)\!=\![ {E}(0),\,\Delta {t} (0)]^{\top}\,\label{eq:space_domain_problem_initial_state},
    \end{align}
\end{subequations}
where ${E}(0)$ and $\Delta {t}(0)$ are the system initial states, which are available in advance. $\tilde{\Psi}(k_s)$ is the terminal cost, which is given by
\begin{equation}
    \begin{aligned}
       & \tilde{\Psi}(k_s)\!=\!W_{E}(\tilde{E}(k_s)- E_{\max}(k_s))^2+W_{\Delta t}(\Delta \tilde{t}(k_s)-\Delta t(0))^2, \label{eq:space_multi_objective_convex_terminal}
    \end{aligned}
\end{equation}
where the $W_{\Delta t}\in\mathbb{R}_{>0}$ term in the terminal cost $\tilde{\Psi}(k_s)$ is imposed to ensure the distance travelled by the ego vehicle is identical to that of the leading vehicle, thereby facilitating the comparison between different methods.
In practice, the $W_{\Delta t}$ term in $\tilde{\Psi}(k_s)$ could be removed or reduced to allow more emphasis on energy economy (more travel time / less energy consumption, since $\Delta \tilde{t}(k_s)$ will tend to $\Delta t_{\max}$) or mobility (less travel time / more energy consumption, since $\Delta \tilde{t}(k_s)$ will tend to $\Delta t_{\min}$), depending on the choice of $W_{E}$ and $W_{F}$.

In the objective function \eqref{eq:space_domain_problem_cost_function}, $\tilde{J}(k)$ is defined as a state-space form of a combination of both the stage cost $\tilde{L}(k)$ \eqref{eq:space_multi_objective_convex_stage} and the terminal cost $\tilde{\Psi}(k_s)$ \eqref{eq:space_multi_objective_convex_terminal}, and it is expressed by
\begin{equation}
    \begin{aligned}
    \label{eq:space_cost_function}
    \tilde{J}(k)&=(\tilde{z}(k)\!-\!\overline{z}(k))^{\!\top}\!Q^{\top}\!Q(\tilde{z}(k)\!-\!\overline{z}(k))\!+\!P\tilde{z}(k)\!+\!\tilde{z}(k)^{\top}\!P^{\top}\!,
    \end{aligned}
\end{equation}
where $\tilde{z}(k)=[\tilde{E}(k),\Delta \tilde{t}(k),F_w(k),\zeta(k)]^{\top}$ and the reference signal $\overline{z}(k)=[E_{\max}(k),\Delta t(0), 0,0]^{\top}$. Furthermore, $\tilde{z}(k)$ can be explicitly expressed as 
\begin{equation}
    \begin{aligned}
    \label{eq:space_cost_function_set}
    \tilde{z}(k)&=C_{z}\tilde{x}(k)+D_{zu}u(k), \\
    C_{z}&\!=\!\begin{bmatrix}
    1&0\\0&1\\0&0\\0&0
    \end{bmatrix}, \, D_{zu}\!\!=\!\!\begin{bmatrix}
    0&0\\0&0\\1&0\\0&1
    \end{bmatrix}. \, 
    \end{aligned}
\end{equation}
The weighting matrices in \eqref{eq:space_cost_function} are defined as
\begin{equation}
    Q(W_{\!\Delta t})\!=\!\begin{bmatrix}
        \sqrt{W_{E}}&0&0&0\\
        0&\sqrt{W_{\Delta t}}&0&0 \\
        0&0&\sqrt{W_{F}a_1}&0\\
        0&0&0&0
    \end{bmatrix}, P\!=\!\begin{bmatrix}
        0\\0\\\frac{1}{2}W_{F}a_2\\\frac{1}{2}W_{\zeta}
    \end{bmatrix},
    \label{eq:space_cost_weighting}
\end{equation}
where the $Q$ matrix is dependent on $W_{\Delta t}$, while $P$ is a fixed matrix.
\subsection{Nominal MPC benchmark}
\label{sec:formulations_benchmark}
This subsection rewrites the convex OCP formulation \eqref{eq:space_domain_problemformulation} into a condensed nominal MPC formulation for the purpose of saving the computational time as well as the memory requirements \cite{Jerez2012}. Moreover, the nominal MPC will be utilised as a benchmark and compared with the proposed REACC scheme in Section \ref{sec:simulation results_robust}. 

Let us now define the following stacked vectors
\begin{equation}
\begin{aligned}
\label{eq:MPC_stack_vectors_nominal}
\mathbf{\tilde{x}}&=[\tilde{x}(0)^{\top},\tilde{x}(1)^{\top},\dots,\tilde{x}(N)^{\top}]^{\top}\in \mathbb{R}^{2(N+1)},\\
\mathbf{u}&=[u(0)^{\top},u(1)^{\top},\dots,u(N-1)^{\top}]^{\top}\in \mathbb{R}^{2N}, \\
\mathbf{C}&=[C(0)^{\top},C(1)^{\top},\dots,C(N-1)^{\top}]^{\top}\in \mathbb{R}^{2N},\\
\mathbf{\xi}&=[\xi(0)^{\top},\xi(1)^{\top},\dots,\xi(N)^{\top}]^{\top},\\
\end{aligned}
\end{equation}
where $N$ is the prediction horizon length of the nominal MPC problem. The symbol $\xi$ stands for 
stacked vectors $\mathbf{\tilde{f}}$, $\mathbf{\underline{f}}$, $\mathbf{\overline{f}}$, $\mathbf{\tilde{z}}$, and $\mathbf{\overline{z}}$, with $\mathbf{\tilde{f}}$, $\mathbf{\underline{f}}$, and $\mathbf{\overline{f}} \in \mathbb{R}^{3(N+1)}$, while $\mathbf{\tilde{z}}$ and $\mathbf{\overline{z}}$ lie in $\mathbb{R}^{4(N+1)}$. As such, the system dynamics \eqref{eq:space_state_function_set} over the prediction horizon $N$ can be rewritten into a condensed formulation as
\begin{equation}
\begin{aligned}
\label{eq:MPC_stack_state_response_nominal}
    \mathbf{\tilde{x}}&=\widetilde{A} \tilde{x}(0)+\widetilde{B}_{u} \mathbf{u} +\widetilde{B}_{c} \mathbf{C}\,,
\end{aligned}
\end{equation}
where $\tilde{x}(0)$ is the initial condition defined in \eqref{eq:space_domain_problem_initial_state}, and $\widetilde{A}$, $\widetilde{B}_{u}$, and $\widetilde{B}_{c}$ are stacked coefficient matrices of $A$, $B_{u}$, and $B_{c}$, respectively. They are readily obtained from iterating the dynamics in \eqref{eq:space_state_function_set} from $k=0$ to $k=N$. 

By repeating the recursive steps in \eqref{eq:space_constraint_function_set} and substituting recursive steps in \eqref{eq:space_state_function_set} to eliminate the $\tilde{x}(k)$ terms, the stacked coefficient matrices $\widetilde{C}_{f}$, $\widetilde{D}_{fu}$, and $\widetilde{D}_{fc}$ are obtained, and hence the corresponding condensed form of the signal response function of constraint \eqref{eq:space_constraint_function_set} can be written after substituting the stacked vectors defined in \eqref{eq:MPC_stack_vectors_nominal} as
\begin{equation}
\begin{aligned}
\label{eq:MPC_stack_constraint_response_nominal}    
\mathbf{\tilde{f}}&=\widetilde{C}_{f}\tilde{x}(0)+\widetilde{D}_{fu}\mathbf{u}+\widetilde{D}_{fc}\mathbf{C}\,,
\end{aligned}
\end{equation}
from which it follows that $\mathbf{\underline{f}}\leq\mathbf{\tilde{f}}\leq\mathbf{\overline{f}}$.

Analogously, the condensed form of the cost response function defined in \eqref{eq:space_cost_function_set} can be expressed by
\begin{equation}
\begin{aligned}
\label{eq:MPC_stack_cost_response_nominal} 
\mathbf{\tilde{z}}&=\widetilde{C}_{z}\tilde{x}(0)+\widetilde{D}_{zu}\mathbf{u}+\widetilde{D}_{zc}\mathbf{C}\,,
\end{aligned}
\end{equation}
where $\widetilde{C}_{z}$, $\widetilde{D}_{zu}$, and $\widetilde{D}_{zc}$ are stacked coefficient matrices after iterating the $\tilde{z}(k)$ equation in \eqref{eq:space_cost_function_set} and substituting the $\tilde{x}(k)$ equation in \eqref{eq:space_state_function_set}. Hence, the stacked formulation of the objective function \eqref{eq:space_cost_function} over a prediction horizon $N$ can be derived by $\mathbf{\tilde{J}}$ with
\begin{equation}\label{eq:space_cost_function_set_nominal_J}
    \begin{aligned}
\mathbf{\tilde{J}}&\!=\!\left(\mathbf{\tilde{z}}-\mathbf{\overline{z}}\right)^{\top}\mathbf{Q}^{\top}\mathbf{Q}\left(\mathbf{\tilde{z}}-\mathbf{\overline{z}} \right)+\mathbf{P}\mathbf{\tilde{z}}+\mathbf{\tilde{z}}^{\top}\mathbf{P}^{\top},
    \end{aligned}
\end{equation}
where $\mathbf{Q}\!=\!\text{diag}\{Q(0),\dots,Q(0),Q(W_{ \Delta t})\}\!\in\!\mathbb{R}^{4(N+1)\times4(N+1)}$,\\
\noindent$\mathbf{P}=[P, \dots, P]\in\mathbb{R}^{1\times 4(N+1)}$ are stacked vectors of ${Q}$ and ${P}$ in \eqref{eq:space_cost_weighting}. Recall the $W_{\Delta t}(\neq0)$ weight is only imposed in the terminal cost as described after \eqref{eq:space_multi_objective_convex_terminal}.

By collecting \eqref{eq:MPC_stack_state_response_nominal}-\eqref{eq:space_cost_function_set_nominal_J}, the condensed nominal MPC can be organised as
\begin{subequations}\label{eq:NominalMPC_final_formula_space}
\begin{align}
     \min\limits_{\mathbf{u}}\quad & \mathbf{\tilde{J}},\\
    \textbf{s.t. } & \mathbf{\tilde{x}}=\widetilde{A} \tilde{x}(0)+\widetilde{B}_{u} \mathbf{u} +\widetilde{B}_{c} \mathbf{C}\,,\\
   &\mathbf{\tilde{f}}=\widetilde{C}_{f}\tilde{x}(0)+\widetilde{D}_{fu}\mathbf{u}+\widetilde{D}_{fc}\mathbf{C}\,,\\
   &\mathbf{\tilde{z}}=\widetilde{C}_{z}\tilde{x}(0)+\widetilde{D}_{zu}\mathbf{u}+\widetilde{D}_{zc}\mathbf{C}\,,\\
   &\mathbf{\underline{f}}\leq\mathbf{\tilde{f}}\leq\mathbf{\overline{f}}\,,\\
    &\begin{bmatrix}\zeta(0)\\\zeta(1)\\\vdots\\ \zeta(N-1)\end{bmatrix}\geq\begin{bmatrix} 1/\sqrt{2\tilde{E}(0)/m}\\ 1/\sqrt{2\tilde{E}(1)/m}\\\vdots\\ 1/\sqrt{2\tilde{E}(N-1)/m}\end{bmatrix}\,,\label{eq:Nominalmpc_space_stack_problem_zeta}\\
    \textbf{given: }& \tilde{x}(0)\!=\![E(0),\,\Delta t (0)]^{\top}.
\end{align}
\end{subequations}
\section{Robust Model Predictive Controller Design for EACC}
\label{sec:rmpc}
This section designs the proposed RMPC controller for solving the convex OCP \eqref{eq:space_domain_problemformulation} in real-time with consideration of two types of disturbances, including the modelling mismatches in the longitudinal dynamics \eqref{eq:space_E_accurate} and the errors involved in the leading vehicle velocity (e.g., caused by communication or sensing) in the dynamics of the time gap \eqref{eq:space_dt_accurate} (proposed REACC scheme). The robust optimal solutions are solved by using SDPR and sLMI methods \cite{Georgiou2023}. 

Based on the nominal ego vehicle longitudinal dynamics \eqref{eq:space_E_tilde}, let us now rewrite the system \eqref{eq:space_E_accurate} in the following form with separated nominal dynamics and additive disturbance
\begin{equation}
\begin{aligned}
     E(k+1)=& E(k) +\bigg((F_w(k)-\frac{2\tilde{f_d}E(k)}{m}\\ 
     &\!\!\!\!-mg\tilde{f_r}\cos\left(\tilde{\theta}(k)\right)-mg\sin\left(\tilde{\theta}(k)\right)+d_E(k)\bigg)\delta s, \label{eq:space_E}
     \end{aligned}  
\end{equation}
where $d_E(k)$ is the disturbance capturing the modelling mismatches existing in both the resistance force coefficients and the road slope angle. In view of \eqref{eq:space_E_accurate}, it holds that
\begin{equation}
\label{eq:disturbance_formula}
\begin{aligned}
d_E(k)
=&\left(\tilde{f}_d\!-\!{f}_d(k)\right)\frac{2E(k)}{m}\\
&\!+\!mg\tilde{f}_r\cos\left(\tilde{\theta}(k)\right)\!+\!mg\sin\left(\tilde{\theta}(k)\right) \\
&\!-\!mg{f}_r(k)\cos\left({\theta}(k)\right)\!-\!mg\sin\left({\theta}(k)\right),
\end{aligned}
\end{equation} 
where $\underline{d}_E\leq d_E(k)\leq \overline{d}_E$ with $\underline{d}_E=\min (d_E(k))$ and $\overline{d}_E=\max(d_E(k))$, which are determined through a conservative consideration of the limits of ${f}_d(k)$, ${f}_r(k)$, $\Delta{\theta}(k)$. 

Furthermore, the utilisation of time gap $\Delta t$ as the system state allows the communication or the sensing error of the leading vehicle velocity to be directly involved in the system dynamics \eqref{eq:space_dt_accurate}, and hence \eqref{eq:space_dt_accurate} can be rewritten as
\begin{equation}
\label{eq:space_dynamics_delta_t}
    \Delta t(k+1)=\Delta t(k)+\left(\frac{1}{\sqrt{2E(k)/m}}-\frac{1}{\tilde{v}_l(k)}+d_{t}(k)\right)\delta s,
\end{equation}
where $\underline{d}_{t}\leq d_{t}(k)\leq \overline{d}_{t}$ is the bounded communication or sensing error of the leading vehicle represented by
\begin{equation}
\label{eq:disturbance_formula_dt}
\begin{aligned}
d_t(k)
=\frac{1}{\tilde{v}_{l}(k)}-\frac{1}{v_{l}(k)},
\end{aligned}
\end{equation} 
where $\underline{d}_{t}\!=\!\text{min}\left(\frac{1}{\tilde{v}_{l}(k)}-\frac{1}{v_{l}(k)} \right)$ and $\overline{d}_{t}\!=\!\text{max}\left(\frac{1}{\tilde{v}_{l}(k)}-\frac{1}{v_{l}(k)} \right)$.
Therefore, based on the nominal state-space equation \eqref{eq:space_state_function_set}, the realistic state-space equation with additive disturbances, after collecting \eqref{eq:space_E} and \eqref{eq:space_dynamics_delta_t}, becomes
\begin{equation}
    {x}(k+1)=A{x}(k)+B_{u}u(k)+B_{c}C(k)+B_{d}d(k), \label{eq:space_state_function_set_real}
\end{equation}
where ${x}(k)\!=\![{E}(k), \Delta {t} (k)]^{\top}$ denotes the realistic state vector,
symbols $A$, $B_{u}$, $B_c$, $u(k)$, and $C(k)$ have the same meaning as in \eqref{eq:space_state_function_set}, 
and furthermore $B_{d}\!=\!\begin{bmatrix}\delta s&0\\0&\delta s\end{bmatrix}$ and $d(k)\!=\!\begin{bmatrix}d_E(k)\\ d_{t}(k)\end{bmatrix}$. In addition, the realistic constraint response function of states and inputs $f(k)$ and the cost response function $z(k)$ are written as
\begin{subequations}
    \begin{align}
    {f}(k)&=C_{f} {x}(k)+D_{fu}u(k)+D_{fd}d(k),\;D_{fd}=\mathbf{0}^{3 \times 2},
    \label{eq:space_constraint_function_set_real}\\
    {z}(k)&=C_{z}{x}(k)+D_{zu}u(k)+D_{zd}d(k),\; D_{zd}=\mathbf{0}^{4\times2}.
    \label{eq:space_cost_function_set_real}
    \end{align}
\end{subequations}
In addition to previously defined stacked vectors in \eqref{eq:MPC_stack_vectors_nominal}, let us now introduce the following extra stacked vectors 
\begin{equation}
\begin{aligned}
\label{eq:MPC_stack_vectors}
\mathbf{x}&=[x(0)^{\top},x(1)^{\top},\dots,x(N)^{\top}]^{\top}\in \mathbb{R}^{2(N+1)},\\
\mathbf{d}&=[d(0)^{\top},d(1)^{\top},\dots,d(N-1)^{\top}]^{\top}\in\mathbb{R}^{2N},\\
\mathbf{\overline{d}}&=[\overline{d}^{\top},\overline{d}^{\top},\dots,\overline{d}^{\top}]^{\top}\in\mathbb{R}^{2N},\\
\mathbf{\underline{d}}&=[\underline{d}^{\top},\underline{d}^{\top},\dots,\underline{d}^{\top}]^{\top}\in\mathbb{R}^{2N},\\
\end{aligned}
\end{equation}
where $N$ is the prediction horizon length of the RMPC problem (which is identical to the nominal MPC to avoid prediction horizon-caused effects on optimisation results). 
$\overline{d}$ and $\underline{d}$ collect disturbance boundaries as $\overline{d}=[\overline{d}_E, \overline{d}_{t}]^{\top}$ and $\underline{d}=[\underline{d}_E,\underline{d}_{t}]^{\top}$, respectively. As such, the condensed system dynamics \eqref{eq:space_state_function_set_real} over the prediction horizon $N$ can be rewritten as
\begin{equation}
\begin{aligned}
\label{eq:MPC_stack_state_response}
    \mathbf{x}&=\widetilde{A} x(0)+\widetilde{B}_{u} \mathbf{u} +\widetilde{B}_{c} \mathbf{C}+\widetilde{B}_{d} \mathbf{d}\,,
\end{aligned}
\end{equation}
$x(0)=\tilde{x}(0)$ is the identical initial condition defined in \eqref{eq:space_domain_problem_initial_state}, and $\widetilde{B}_{d}$ is the stacked coefficient matrix of $B_{d}$, which can be obtained similarly as matrices $\widetilde{A}$, $\widetilde{B}_{u}$, and $\widetilde{B}_{c}$ obtained in \eqref{eq:MPC_stack_state_response_nominal}.

By respectively repeating the recursive steps in \eqref{eq:space_constraint_function_set_real} and \eqref{eq:space_cost_function_set_real} from $k=0$ to $k=N$, the corresponding condensed format of the constraint and the cost response functions can be written as
\begin{subequations}
    \begin{align}    \mathbf{f}&=\widetilde{C}_{f}x(0)+\widetilde{D}_{fu}\mathbf{u}+\widetilde{D}_{fc}\mathbf{C}+\widetilde{D}_{fd}\mathbf{d}\,,\label{eq:MPC_stack_constraint_response} \\
    \mathbf{z}&=\widetilde{C}_{z}x(0)+\widetilde{D}_{zu}\mathbf{u}+\widetilde{D}_{zc}\mathbf{C}+\widetilde{D}_{zd}\mathbf{d}\,,\label{eq:MPC_stack_cost_response}
    \end{align}
\end{subequations}
where 
stacked matrices $\widetilde{D}_{fd}$ and $\widetilde{D}_{zd}$ can be analogously obtained after recursively iterating \eqref{eq:space_constraint_function_set_real} and \eqref{eq:space_cost_function_set_real}, respectively,  as in previous steps. Moreover, the condensed constraint response function \eqref{eq:MPC_stack_constraint_response} satisfies the stacked constraint $\mathbf{\underline{f}}\leq\mathbf{f}\leq\mathbf{\overline{f}}$.

Therefore, similarly to the nominal condensed objective function $\mathbf{\tilde{J}}$ in \eqref{eq:space_cost_function_set_nominal_J}, the condensed formulation of the objective function considering disturbances over the prediction horizon $N$ can be derived by $\mathbf{J}$ in \eqref{eq:space_cost_function_set_robust_J0} followed by substituting \eqref{eq:MPC_stack_cost_response} for $\mathbf{z}$ to obtain \eqref{eq:space_cost_function_set_robust_J1}, where
\begin{subequations}
    \begin{align}
\mathbf{J}&\!=\!\left(\mathbf{{z}}-\mathbf{\overline{z}}\right)^{\top}\mathbf{Q}^{\top}\mathbf{Q}\left(\mathbf{{z}}-\mathbf{\overline{z}} \right)+\mathbf{P}\mathbf{{z}}+\mathbf{{z}}^{\top}\mathbf{P}^{\top}, \label{eq:space_cost_function_set_robust_J0}\\
&\!=\!(\widetilde{C}_{z}x(0)\!+\!\widetilde{D}_{zu}\mathbf{u}\!+\!\widetilde{D}_{zc}\mathbf{C}\!+\!\widetilde{D}_{zd}\mathbf{d}\!-\!\mathbf{\overline{z}})^{\top}\mathbf{Q}^{\top} \nonumber\\
    &\qquad \mathbf{Q} (\widetilde{C}_{z}x(0)\!+\!\widetilde{D}_{zu}\mathbf{u}\!+\!\widetilde{D}_{zc}\mathbf{C}\!+\!\widetilde{D}_{zd}\mathbf{d}\!-\!\mathbf{\overline{z}}) \nonumber\\
    &\qquad +\mathbf{P}(\widetilde{C}_{z}x(0)\!+\!\widetilde{D}_{zu}\mathbf{u}\!+\!\widetilde{D}_{zc}\mathbf{C}\!+\!\widetilde{D}_{zd}\mathbf{d}) \label{eq:space_cost_function_set_robust_J1} \\
    &\qquad +(\widetilde{C}_{z}x(0)\!+\!\widetilde{D}_{zu}\mathbf{u}\!+\!\widetilde{D}_{zc}\mathbf{C}\!+\!\widetilde{D}_{zd}\mathbf{d})^{\top}\mathbf{P}^{\top}. \nonumber    
    \end{align}
\end{subequations}

Next, a new auxiliary variable, $\overline{\gamma}$, is introduced to represent the upper bound of stacked objective functions $\mathbf{J}$ such that
\begin{equation}
    \begin{aligned}
    \mathbf{J}- \overline{\gamma}\leq 0, \forall \mathbf{d}:\mathbf{\underline{d}} \leq \mathbf{d} \leq \mathbf{\overline{d}} .\label{eq:RMPC_gamma}
    \end{aligned}
\end{equation}
Furthermore, we use an SDPR procedure to turn \eqref{eq:RMPC_gamma} into a semi-definite program. It can be verified using \eqref{eq:space_cost_function_set_robust_J1} that $\mathbf{J}-\overline{\gamma}$ can be written as 
\begin{equation}
    \begin{aligned}
    \mathbf{J}- \overline{\gamma}=-(\mathbf{d}-\underline{\mathbf{d}})^{\top}D(\mathbf{\overline{d}}-\mathbf{d})-\begin{bmatrix}\mathbf{d}^{\top}&1\end{bmatrix}L(\mathbf{u},D,
    \overline{\gamma})\begin{bmatrix}\mathbf{d}\\1\end{bmatrix}, \label{eq:RMPC_gamma_LHS}
    \end{aligned}
\end{equation}
where $D\succeq 0$ with $D \in \mathbb{D}^{2N}$ is a positive semi-definite diagonal matrix, and $L(\mathbf{u},D,\overline{\gamma})$ is a matrix dependent on $\mathbf{u}$, $D$, and $\overline{\gamma}$, such that 
\begin{equation}
\begin{aligned}
&L(\mathbf{u},D,\overline{\gamma})=\\
    &\begin{bmatrix}
    -\widetilde{D}_{zd}^{\top}\mathbf{Q}^{\!\top}\!\mathbf{Q}\widetilde{D}_{zd}\!+\!
    D & -D(\underline{\mathbf{d}}+\overline{\mathbf{d}})/{2}-bd \\
    * & \underline{\mathbf{d}}^{\!\top}\!D\overline{\mathbf{d}}\!-\!cd\!-\!\mathbf{u}^{\!\top}\!\widetilde{D}_{zu}^{\top}\!\mathbf{Q}^{\!\top}\!\mathbf{Q}\widetilde{D}_{zu}\mathbf{u}\!+\!\overline{\gamma}\end{bmatrix}, \label{eq:RMPC_L_objective} 
    \end{aligned} 
\end{equation}   
with $*$ denoting the symmetry element of the corresponding matrix and   
\begin{equation*}
\begin{aligned}
&bd\!=\!\widetilde{D}_{zd}^{\top}\mathbf{Q}^{\top}\mathbf{Q}\widetilde{C}_{z}x(0)\!+\!\widetilde{D}_{zd}^{\top}\mathbf{Q}^{\top}\mathbf{Q}\widetilde{D}_{zu}\mathbf{u}+\\
&\qquad\widetilde{D}_{zd}^{\top}\mathbf{Q}^{\top}\mathbf{Q}\widetilde{D}_{zc}\mathbf{C}\!-\!\widetilde{D}_{zd}^{\top}\mathbf{Q}^{\top}\mathbf{Q}\mathbf{\overline{z}}\!+\!\widetilde{D}_{zd}^{\top}\mathbf{P}^{\top}\!,\\
&cd\!=\!x(0)^{\top}\widetilde{C}_{z}^{\top}\mathbf{Q}^{\top}\mathbf{Q}\widetilde{C}_{z}x(0)\!+\!x(0)^{\top}\widetilde{C}_{z}^{\top}\mathbf{Q}^{\top}\mathbf{Q}\widetilde{D}_{zu}\mathbf{u}+\\
&\qquad x(0)^{\top}\widetilde{C}_{z}^{\top}\mathbf{Q}^{\top}\mathbf{Q}\widetilde{D}_{zc}\mathbf{C}\!+\!\mathbf{u}^{\top}\widetilde{D}_{zu}^{\top}\mathbf{Q}^{\top}\mathbf{Q}\widetilde{C}_{z}x(0)+\\
&\qquad\mathbf{u}^{\top}\widetilde{D}_{zu}^{\top}\mathbf{Q}^{\top}\mathbf{Q}\widetilde{D}_{zc}\mathbf{C}\!+\!\mathbf{C}^{\top}\widetilde{D}_{zc}^{\top}\mathbf{Q}^{\top}\mathbf{Q}\widetilde{C}_{z}x(0)+\\
&\qquad\mathbf{C}^{\top}\widetilde{D}_{zc}^{\top}\mathbf{Q}^{\top}\mathbf{Q}\widetilde{D}_{zu}\mathbf{u}\!+\!\mathbf{C}^{\top}\widetilde{D}_{zc}^{\top}\mathbf{Q}^{\top}\mathbf{Q}\widetilde{D}_{zc}\mathbf{C}-\\
 &\qquad x(0)^{\top}\widetilde{C}_{z}^{\top}\mathbf{Q}^{\top}\mathbf{Q}\mathbf{\overline{z}}\!-\!\mathbf{u}^{\top}\widetilde{D}_{zu}^{\top}\mathbf{Q}^{\top}\mathbf{Q}\mathbf{\overline{z}}\!-\!\mathbf{C}^{\top}\widetilde{D}_{zc}^{\top}\mathbf{Q}^{\top}\mathbf{Q}\mathbf{\overline{z}}-\\
&\qquad\mathbf{\overline{z}}^{\top}\mathbf{Q}^{\top}\mathbf{Q}\widetilde{C}_{z}x(0)\!-\!\mathbf{\overline{z}}^{\top}\mathbf{Q}^{\top}\mathbf{Q}\widetilde{D}_{zu}\mathbf{u}\!-\!\mathbf{\overline{z}}^{\top}\mathbf{Q}^{\top}\mathbf{Q}\widetilde{D}_{zc}\mathbf{C}+\\
&\qquad\mathbf{\overline{z}}^{\top}\mathbf{Q}^{\top}\mathbf{Q}\mathbf{\overline{z}}\!+\!\mathbf{P}(\widetilde{C}_{z}x(0)+\widetilde{D}_{zu}\mathbf{u}\!+\!\widetilde{D}_{zc}\mathbf{C})+\\
&\qquad (\widetilde{C}_{z}x(0)\!+\!\widetilde{D}_{zu}\mathbf{u}\!+\!\widetilde{D}_{zc}\mathbf{C})^{\top}\mathbf{P}^{\top}. \nonumber\\
\end{aligned}
\end{equation*}
An inspection of \eqref{eq:RMPC_gamma_LHS} verifies that \eqref{eq:RMPC_gamma} is satisfied if $D\succeq 0$ with $D \in \mathbb{D}^{2N}$ and if the following LMI is achieved
\begin{equation}
    \begin{aligned}
    L(\mathbf{u},D,\overline{\gamma})\succeq 0. \label{eq:RMPC_gamma_LMI1_time}
    \end{aligned}
\end{equation}
However, the quadratic term $\mathbf{u}^{\!\top}\!\widetilde{D}_{zu}^{\top}\!\mathbf{Q}^{\!\top}\!\mathbf{Q}\widetilde{D}_{zu}\mathbf{u}$ in the 2,2 entry of $L(\mathbf{u},D,\overline{\gamma})$ in \eqref{eq:RMPC_L_objective} makes the matrix $L(\mathbf{u},D,\overline{\gamma})$ nonlinear. In order to satisfy the linearity requirement of the LMI optimisation, Schur complement is adopted here to eliminate the nonlinearity, leading to the redefinition of 
\begin{equation}
\begin{aligned}
    &L(\mathbf{u},D,\overline{\gamma})
    =\\
    &\begin{bmatrix}
    -\widetilde{D}_{zd}^{\top}\mathbf{Q}^{\top}\mathbf{Q}\widetilde{D}_{zd}+
    D \!&\! -D(\underline{\mathbf{d}}+\overline{\mathbf{d}})/{2}\!-\!bd
    \!&\! 0\\
    * \!&\! \underline{\mathbf{d}}^{\top}D\overline{\mathbf{d}}\!-\!cd+\overline{\gamma}&\mathbf{u}^{\top}\widetilde{D}_{zu}^{\top}\mathbf{Q}^{\top}\\
    *\!&\!*\!&\quad{I}^{4(N+1)}\end{bmatrix}.\label{eq:RMPC_L_objective_linearised}
\end{aligned}
\end{equation}

In view of \eqref{eq:MPC_stack_constraint_response}, the stacked inequality constraints $\mathbf{\underline{f}}\leq\mathbf{f}\leq\mathbf{\overline{f}}$ can be expanded as
\begin{equation}
    \mathbf{\underline{f}}\leq   \widetilde{C}_{f}x(0)+\widetilde{D}_{fu}\mathbf{u}+\widetilde{D}_{fc}\mathbf{C}+\widetilde{D}_{fd}\mathbf{d}\,
    \leq \mathbf{\overline{f}}, \label{eq:Robust_space_constraint_upper0}
\end{equation}
which normally requires $3(N+1)$ LMI constraints for the upper and lower boundaries, respectively \cite{Yu2022ECC}. To reduce the computational burden, the sLMI approach which combines all $6(N+1)$ LMIs in \eqref{eq:Robust_space_constraint_upper0} into a single LMI is presented next. First, \eqref{eq:Robust_space_constraint_upper0} is written as
\begin{equation}
    \begin{aligned}
        \begin{bmatrix}
            ~~{I}^{3(N+1)}\\-{I}^{3(N+1)}
        \end{bmatrix}\mathbf{f} \leq \begin{bmatrix}
            ~~\mathbf{\overline{f}}\\\mathbf{-\underline{f}}
        \end{bmatrix}=:\mathbf{\overline{f}}^*
        \label{eq:Robust_constraint_single_LMI}.
    \end{aligned}
\end{equation}
To simplify the notation, let $\mathbf{I}^*=\begin{bmatrix}~~{I}^{3(N+1)}\\-{I}^{3(N+1)}\end{bmatrix}$.
By defining a new variable $\tilde{\mathbf{f}}\in  \mathbb{R}^{6(N+1)}$, it is desirable to satisfy the single condition
\begin{equation}
    \begin{aligned}
        \tilde{\mathbf{f}}:=\mathbf{\overline{f}}^*-\mathbf{I}^*\mathbf{f}\geq 0 , \label{eq:single constraint}
    \end{aligned}
\end{equation}
such that constraints \eqref{eq:Robust_space_constraint_upper0} are satisfied. From Theorem 3 of \cite{Georgiou2023}, let 
$e\in \mathbb{R}^{6(N+1)} $ be the vector of ones, then 
$ \tilde{\mathbf{f}}\geq 0$ if there exist $\mu \in \mathbb{R}$ and  $M \in \mathbb{D}^{6(N+1)}$ such that
\begin{equation}
    \begin{aligned}
       \mathscr{L}= \begin{bmatrix}
           2\mu & (\tilde{\mathbf{f}}-Me-e\mu)^{\top}\\
           * & M+M^\top
       \end{bmatrix} \succeq 0. \label{eq:scr_L}
    \end{aligned}
\end{equation}
By substituting \eqref{eq:single constraint} for $\tilde{\mathbf{f}}$ and \eqref{eq:MPC_stack_constraint_response} for $\mathbf{f}$ into \eqref{eq:scr_L}, yields
\begin{equation}
    \begin{aligned}
       \!\!\begin{bmatrix}
           2\mu \!\! & \!\!\left(\mathbf{\overline{f}}^*\!\!-\!\mathbf{I}^*\!\!\left(\!\!\widetilde{C}_{\!f}\!x(0)\!+\!\widetilde{D}_{\!fu}\!\mathbf{u}\!+\!\widetilde{D}_{\!fc}\!\mathbf{C}\!+\!\widetilde{D}_{\!fd}\mathbf{d}\!\! \right)\!\!-\!\left(Me\!+\!e\mu \right)\!\!\right)^{\!\!\top}\\
           * & M+M^\top
       \end{bmatrix} \!\!\!\succeq \!0.  \label{eq:scr_L_substituted2}
    \end{aligned}
\end{equation}
Applying the Schur complement to transform the left-hand-side of $\eqref{eq:scr_L_substituted2}$ into a scalar and followed by an SDPR procedure to remove the disturbance term $\mathbf{d}$, one can obtain a nonlinear matrix inequality constraint
\begin{equation}
    L_{sLMI}(\tilde{D}, \mu, M, \mathbf{u}) \succeq 0, \label{eq:scr_L_slmi}
\end{equation}
where $0\preceq\tilde{D}\in \mathbb{D}^{2N} $ is a new slack variable. To remove the nonlinearity included, the Schur complement is applied again such that a linear $L_{sLMI}(\tilde{D}, \mu, M, \mathbf{u})$ is determined as 
\begin{equation}
    \begin{aligned}
        &L_{sLMI}(\tilde{D}, \mu, M, \mathbf{u})=\\
        &\!\!\!\!\!\begin{bmatrix}
            \tilde{D}\!\!& -\tilde{D}\overline{\mathbf{d}} & \widetilde{D}_{fd}^{\top}\mathbf{I}^{*\top}\\
            * \!\!&2 \mu \!\!+\!\! \underline{\mathbf{d}}^{\top}\tilde{D}\overline{\mathbf{d}} \!\!&\!\! \left(\!\!- \mathbf{\overline{f}}^*\!\!\!+\!\!\mathbf{I}^*\!\!\left(\widetilde{C}_{f}x(0)\!\!+\!\!\widetilde{D}_{fc}\mathbf{C}\!\!+\!\!\widetilde{D}_{fu}\mathbf{u}\!\right)\!\!+\!\!\left(Me\!\!+\!\!e\mu \right)\!\!\right)^{\!\!\!\top}\\
            *\!\!&*&  \!\!\!\!\left(\!\!M\!\!+\!\!M^{\!\!\top}\!\!\right)\!\!
        \end{bmatrix}\!. \label{eq:L_afterSDPR_linear3}
    \end{aligned}
\end{equation}
\begin{remark}
\label{rem:appendix_sLMI_proof}
Detailed derivation steps from LMI  \eqref{eq:scr_L_substituted2} to matrix  \eqref{eq:L_afterSDPR_linear3} can be referred to Appendix \ref{app:sLMI Proof}.
\end{remark}

Therefore, if the single LMI \eqref{eq:scr_L_slmi} of \eqref{eq:L_afterSDPR_linear3} is satisfied, all constraints in \eqref{eq:Robust_space_constraint_upper0} are satisfied.


To summarise, the RMPC formulation of the optimisation problem under the space-domain scheme, after applying SDPR, Schur complement and sLMI, can be organised as
\begin{subequations}\label{eq:RobustMPC_final_formula_space}
\begin{align}
     \min\limits_{\mathbf{u}}\quad & \overline{\gamma},\\
    \textbf{s.t. } & L(\mathbf{u},D,\overline{\gamma})\succeq 0,\\
    & L_{sLMI}(\tilde{D}, \mu, M, \mathbf{u}) \succeq 0, \\
    &\begin{bmatrix}\zeta(0)\\\zeta(1)\\\vdots\\ \zeta(N-1)\end{bmatrix}\geq\begin{bmatrix} 1/\sqrt{2E(0)/m}\\ 1/\sqrt{2E(1)/m}\\\vdots\\ 1/\sqrt{2E(N-1)/m}\end{bmatrix}\,,\label{eq:mpc_space_stack_problem_zeta}\\
    \textbf{given: }& x(0)\!=\![E(0),\,\Delta t (0)]^{\top},\\
    & 0\preceq D \in \mathbb{D}^{2N}, \,\\
    & 0\preceq\tilde{D}\in \mathbb{D}^{2N}.
\end{align}
\end{subequations}
\section{Simulation Results}
\label{sec:simulation results}
The evaluation of the proposed space-domain REACC method is fourfold: 1) the robustness of the RMPC method is investigated and compared with a nominal benchmark MPC method (described in Section \ref{sec:formulations_benchmark}) given the same initial conditions and disturbances in simulations; 
2) a comprehensive comparison is conducted between space-domain (denoted by REACC) and time-domain formulations described in \cite{Yu2023ICM} in terms of energy consumption and passenger comfort using the same RMPC method in both domains under identical initial conditions and disturbances; 
3) the proposed REACC is further compared with a benchmark method using a CDFS that targets a fixed inter-vehicular distance gap in terms of energy efficiency;
4) the computational efficiency of the sLMI-based REACC method is examined by evaluating the average running time required for each iteration.
The numerical simulations are tested in the Matlab environment using the optimisation toolbox Yalmip \cite{Lofberg2004} with MOSEK solver \cite{mosek} on a 2.3~GHz quad-core Intel Core i5 with an 8~GB of 2133~MHz LPDDR3 onboard memory. Before presenting the numerical examples, the velocity profiles of the leading vehicle and the definition of the disturbances are given.
\subsection{Simulation setup}
\begin{figure}[t!]
\centering
\includegraphics[width=\columnwidth]{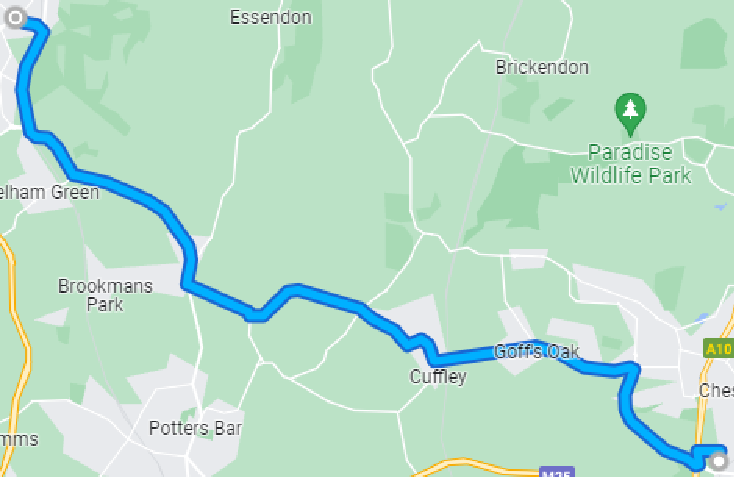}
  \setlength{\belowcaptionskip}{-5pt}
  \caption{$18.7$~km route in London UK selected for the velocity profile of the leading vehicle in the numerical simulations. (https://goo.gl/maps/2CTCW7smdCkGCsKv5)}
  \label{fig:Rural Road}
\end{figure}
\begin{figure*}[t!]
\centering
\includegraphics[width=\textwidth]{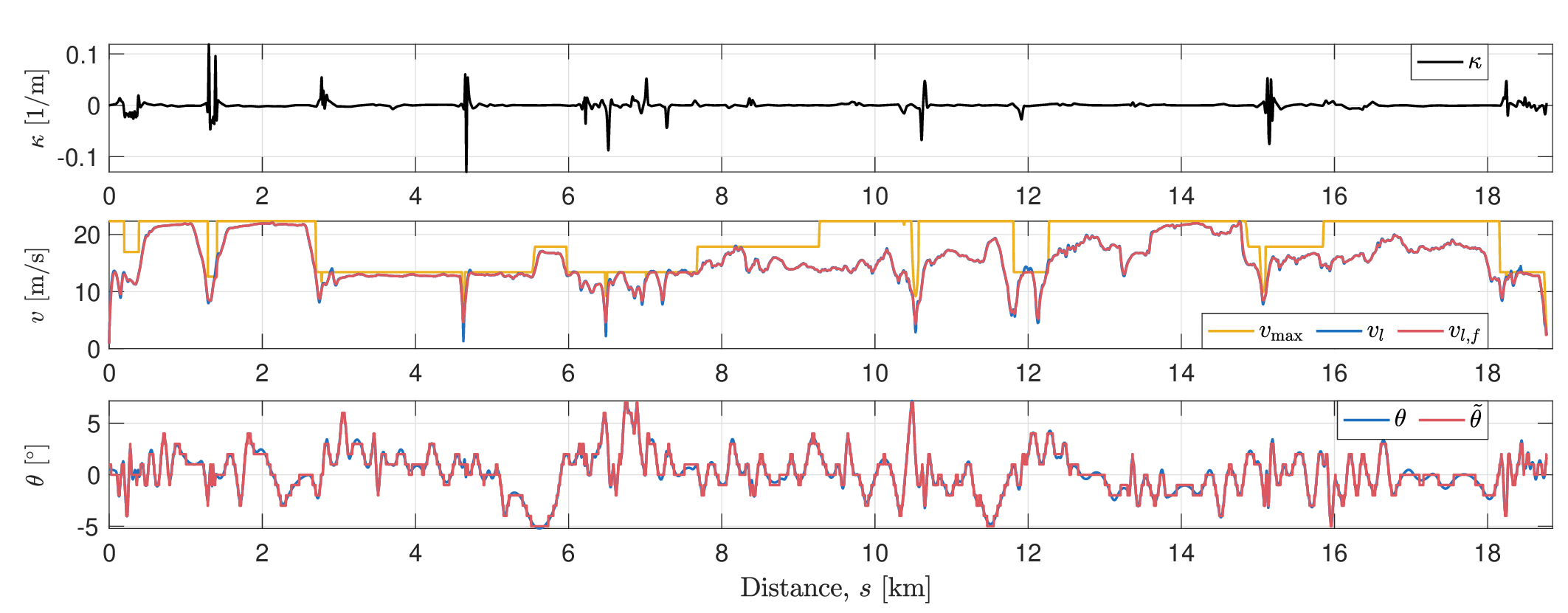}\\[-2ex]
\setlength{\belowcaptionskip}{-15pt}
\caption{Detailed road information of the London UK route with: a) actual road curvature (top), b) actual leading vehicle velocity $v_l$, `filtered' leading vehicle velocity $v_{l,f}$, and combined speed limit $v_{\max}$ (middle), and c) actual and nominal road slope angles, $\theta$ and $\tilde{\theta}$, respectively (bottom).}
\label{fig:Road_profile}
\end{figure*}

The leading vehicle in the following simulations is operated on an experimental route, of which the data is collected on a real-world route in London UK as shown in Fig. \ref{fig:Rural Road}. The road profile data and the traffic constraint of the selected test route are collected based on \emph{Google Maps} and are shown in Fig. \ref{fig:Road_profile}, which includes a plot on the road curvature profile (top subplot), a plot on the leading vehicle velocity, its prediction, and the combined speed limit (middle subplot), and a plot on the road slope angle (bottom subplot). 
Specifically, the top subplot of Fig.~\ref{fig:Road_profile} shows the actual curvature $\kappa$ of the test route. The middle subplot contains three velocity profiles represented by different colours. The yellow curve denotes the combined speed limit of this selected route, $v_{\max}(v_{leg}, \kappa$), which is dependent on the legal speed limit as well as the cornering speed limit which is further dependent on the road curvature (calculated by \eqref{eq:nonlinear_v_constraint}). 
The blue trajectory represents the actual velocity profile of the leading vehicle (i.e., ${v}_{l}$), which follows an experimental velocity profile collected by a human-driven passenger car on the real route. It can be observed that this non-optimised velocity profile could involve aggressive manoeuvres (e.g., at $s$=4.6\,km) and violations of the speed limit (e.g., at $s$=12\,km).
The implementation of MPC demands the prediction of $\Delta t$ (by \eqref{eq:space_delta_t_tilde} or \eqref{eq:space_dynamics_delta_t}) within the control horizon, which in turn requires a prediction of the leading vehicle velocity, $\tilde{v}_l$. 
To this end, we introduce the red curve ($v_{l,f}$) that is obtained by passing the real velocity profile (the blue curve) through a low-pass filter to remove sharp changes in speed, which can be understood as unexpected events and high-frequency noises from communication or sensing. Furthermore, the filtered speed profile is capped by the legal speed limit profile assuming the predicted velocity profile obeys the legal speed limit. 
Then, the prediction of the leading vehicle $\tilde{v}_{l}(k{:}k{+}N{-}1)$ available to the ego vehicle is generated by setting $\tilde{v}_{l}(k{:}k{+}N{-}1){=}v_{l,f}(k{:}k{+}N{-}1)$,~$\forall k \!\in\! [0,k_s]$) to emulate non-optimal behaviours and inevitable communication or sensing error of the leading vehicle velocity. The resulting velocity error further results in the disturbance on the time gap state, $d_{t}$, 
(determined by \eqref{eq:disturbance_formula_dt}). 

The bottom subplot of Fig.~\ref{fig:Road_profile} illustrates the slope angle profiles of the road, where the blue curve denotes the actual road slope angle trajectory $\theta$ collected from the latest \emph{Google Maps} database, with $\theta(k)\!\in\![\underline{\theta}, \overline{\theta}]$. The nominal (available to the ego vehicle) slope angle profile, $\tilde{\theta}$ (red curve in the same subplot), is obtained after rounding $\theta$ to integers in degrees. The difference between the actual and nominal slope angle leads to the associated modelling mismatch $\Delta \theta$ in \eqref{eq:space_define_modelling_mismatches}. 

In addition to the modelling mismatch on the road slope angle as well as the communication or the sensing error on the velocity of the leading vehicle, the disturbances considered in this work also come from modelling mismatches on the air-drag and the tyre-rolling resistance coefficients. The practical air-drag resistance coefficients utilised in this work are obtained according to a fitted lookup table associated with the real-time inter-vehicular distance gap \cite{Lopes2019}, which are also used to provide $\underline{f}_d$ and $\overline{f}_d$. The actual tyre-rolling resistance coefficients adopted in the simulations are randomly generated within the specified bounds of the coefficient, $\underline{f}_r$ and $\overline{f}_r$, with a uniform distribution. The nominal coefficients of the air-drag and tyre-rolling resistance, $\tilde{f}_d$ and $\tilde{f}_r$ respectively, are determined by the middle points of the associated bounds. 
Furthermore, the overall disturbance on the ego vehicle kinetic energy $d_{E}$ and its limits are obtained by merging all modelling mismatches on $f_d$, $f_r$, and $\theta$, and their limits, respectively, by \eqref{eq:disturbance_formula}, with $d_{E}(k)\!\in\![\underline{d}_{E},\overline{d}_{E}]$. Additionally, the sampling distance interval is set to $\delta s{=}3$\,m. 

The limits of disturbances discussed as well as other main characteristic parameters of the ego vehicle model are summarised in Table.~\ref{tab:model_specification_parameters2}.
\begin{table}[t!]
    \centering
        \caption{\textsc{Vehicle Parameters and Road Characteristics.}}
    \label{tab:model_specification_parameters2}
    \begin{tabular*}{1\columnwidth}{c @{\extracolsep{\fill}} c@{\extracolsep{\fill}}c}
        \hline
        \hline
          Description & Symbols & Values \\
         \hline
         Acceleration of gravity& $g$ & 9.81 $\mathrm{m/s^2}$ \\

         Ego vehicle mass& $m$ & 1200~kg\\
         
         \begin{tabular}{c@{}c@{}}Nominal air-drag 
         coefficient \end{tabular}& $\tilde{f}_{d}$ & $0.34$~kg/m\\

         \begin{tabular}{c@{}c@{}}Nominal tyre-rolling \\resistance coefficient \end{tabular}& $\tilde{f}_{r}$ & $0.01$ \\
         
         Ego vehicle initial velocity& $v_0$ & 0.9108~m/s\\

         Initial time gap  & $\Delta t(0)$ & 3~s\\
         
         Acceleration limits & $a_{x,max}$/$a_{y,max}$ &  9.81 $\mathrm{m/s^2}$\\
         
         Min velocity limit & $v_{\min}$ & 0.1~m/s   \\ 
         
         Min/max time gap& $\Delta t_{\min}$/$\Delta t_{\max}$ & 1/8~s\\
         
         \begin{tabular}{c@{}c@{}}Limits of air-drag 
         coefficient \end{tabular} & $\underline{f}_d$/$\overline{f}_d$ & $0.296/0.380$~kg/m \\
         
         \begin{tabular}{c@{}c@{}}Limits of tyre-rolling \\resistance coefficient \end{tabular} & $\underline{f}_r$/$\overline{f}_r$ & $0.008/0.012$ \\

         Limits of road slope & $\underline{\theta}$/$\overline{\theta}$ &
         $-5.22^{\circ}/7.17^{\circ}$\\ 
         
         Limits of road slope mismatch & $\underline{\Delta \theta}$/$\overline{\Delta \theta}$ & $-0.5^{\circ}/0.5^{\circ}$\\ 
         \begin{tabular}{@{}c@{}}Limits of disturbance \\ on kinetic energy \end{tabular} & $\underline{d}_{E}$/$\overline{d}_{E}$ & $
         -146.28/148.20$~N\\
        \begin{tabular}{@{}c@{}}Limits of disturbance \\ on time gap \end{tabular}   & $\underline{d}_{t}$/$\overline{d}_{t}$ & $-0.58/0.11$~s/m\\
         \hline
         \hline
    \end{tabular*}
\end{table}
\subsection{Comparisons between nominal MPC and RMPC} \label{sec:simulation results_robust}
\begin{figure}[t!]
\centering
\includegraphics[width=\columnwidth]{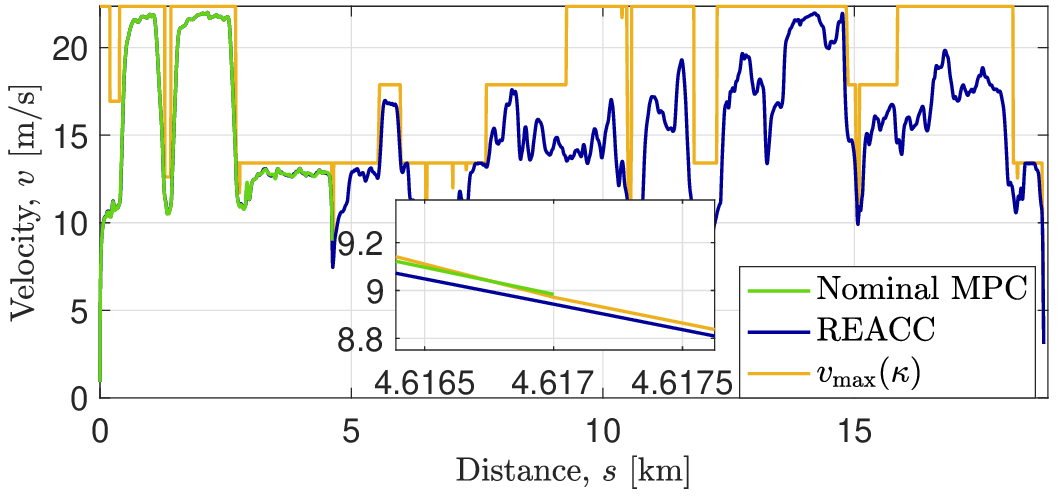}\\[-1.5ex]
\caption{A comparison of the ego vehicle speed trajectories between the REACC method and the nominal MPC under the same initial conditions and simulation disturbances.}
\label{fig:Result_robust_compare}
\end{figure}

A comparison of the ego vehicle speed trajectories between the REACC proposed in \eqref{eq:RobustMPC_final_formula_space} and the nominal MPC benchmark formulated in \eqref{eq:NominalMPC_final_formula_space} is illustrated in Fig. \ref{fig:Result_robust_compare}. After simulating with identical disturbances in both the nominal MPC benchmark as well as the proposed REACC scheme, it can be observed that when the travelled distance reaches $4.617$~km in the zoomed-in box of Fig. \ref{fig:Result_robust_compare}, the velocity trajectory of the nominal MPC violates the speed limit constraint, which leads to infeasible solutions. In contrast, the robust controller (REACC) can always satisfy the velocity constraint with feasible solutions, which verifies the robustness of the RMPC. 
\subsection{Comparison between time- and space-domain schemes}
\label{sec:comparison_time_space}
In this section, the performance of the space-domain formulation is investigated by comparing the proposed space-domain formulated REACC scheme against the results yielded by a benchmark scheme using the same RMPC but formulated in the time-domain utilising an energy consumption model based on the $L^{2}$-norm of the acceleration~\cite{Yu2023ICM}, under the same initial conditions and disturbances. 
The sampling interval of the time-domain scheme is chosen as $0.2$~s, which guarantees the equivalence of the total number of samples of the two domain schemes.
For the sake of a fair comparison among the following numerical examples, the weights in the cost functions of both time- and space-domain MPC schemes are finely adjusted for a small number of iterations at the end of the simulation to ensure negligible differences in terms of the terminal speed (the terminal speeds of the ego vehicles in both domains and the final speed of the leading vehicle are all the same).
As such, the involvement of the kinetic energy change can be excluded and the total energy consumption during a driving cycle can be directly compared as the battery energy usage, $E_{b}\!=\!\sum_{k=0}^{k_s-1}P_{b}(F_{t}^*(k),v^*(k))\frac{\delta s}{v^*(k)}$, which is a part of the original multi-objective cost function in \eqref{eq:space_multi_objective0}. $F_{t}^*(k)$ is the optimal input force on the wheels and $v^*(k)$ is the optimal speed determined by the utilised control methods.

\begin{figure}[t!]
\centering
\includegraphics[width=\columnwidth]{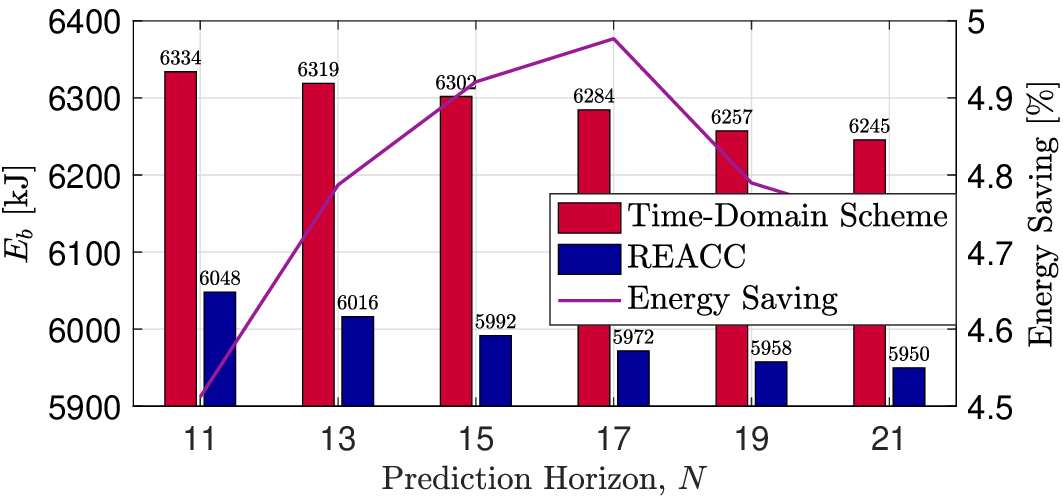}\\[-1ex]
\caption{Comparisons of the battery energy consumption $(E_{b})$ of the REACC and time-domain scheme using the same RMPC, for different prediction horizon lengths.}
\label{fig:Result_robust_energy}
\end{figure}

\begin{figure}[t!]
\centering
\includegraphics[width=\columnwidth]{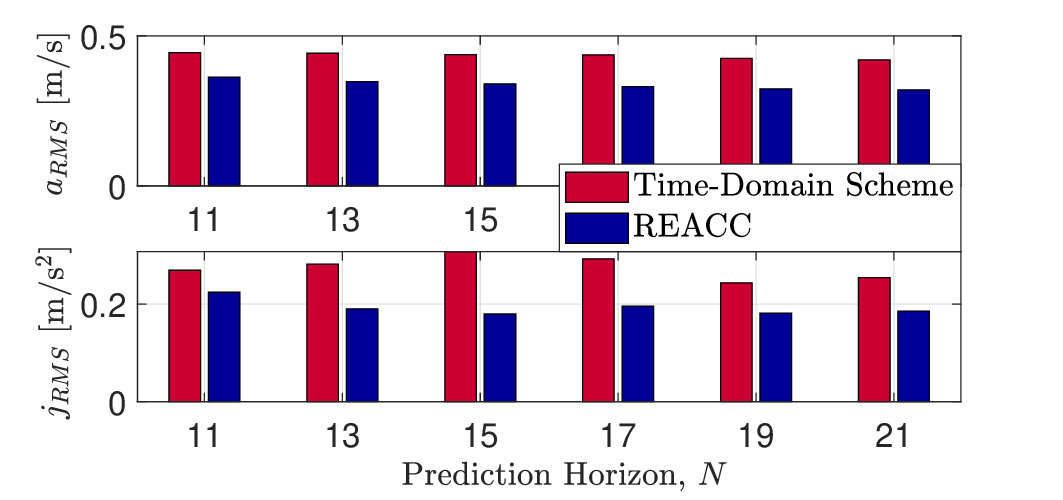}\\[-1ex]
\caption{Comparisons of the RMS acceleration and RMS jerk throughout the mission of the REACC and time-domain scheme using the same RMPC, for different prediction horizon lengths.
}
\label{fig:Result_domain_comfort}
\end{figure}

In Fig. \ref{fig:Result_robust_energy}, the comparison of the battery energy consumption of the space-domain (denoted by REACC) and the time-domain schemes using the same RMPC is presented.
As it can be seen for all cases in Fig.~\ref{fig:Result_robust_energy}, the REACC scheme can save around 4.8\% battery energy consumption as compared with the results of the time-domain formulated benchmark. In addition, the largest improvement of roughly 5\% can be found when the prediction horizon is $17$. 
Further investigation of the driving comfort verifies the finding of the energy saving of the REACC strategy. As shown in Fig. \ref{fig:Result_domain_comfort}, the driving comfort is evaluated by the root-mean-square (RMS) of acceleration and jerk $\left(j\!=\!\frac{d^2v}{dt^2}\right)$ of the ego vehicle. The proposed method achieves lower values of both indexes against the time-domain benchmark for all prediction horizon length choices. 
The findings in both Fig.~\ref{fig:Result_robust_energy} and Fig. \ref{fig:Result_domain_comfort} can be understood that as compared to the time-domain formulated benchmark method, the REACC strategy, in addition to the highly accurate and convexified powertrain fitting model embedded into the cost function, which is exclusively enabled in the space-domain formulation, tends to avoid large accelerations and achieve a smooth driving profile, hence is expected to be more energy efficient in terms of powertrain operation.
\subsection{Comparison between CDFS and proposed REACC schemes}
\begin{figure}[t!]
\centering
\includegraphics[width=\columnwidth]{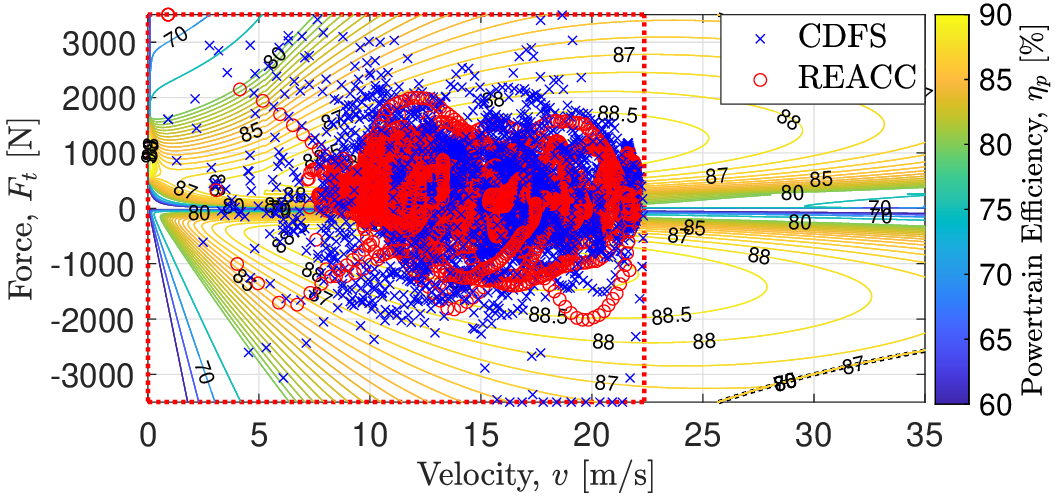}\\[-1ex]
\caption{Operating points of REACC ($N$=11) and CDFS schemes.}
\label{fig:Result_operation_point_compare}
\end{figure}

This section investigates the energy economy of the proposed REACC against a benchmark using a CDFS scheme by which the velocity of the ego vehicle is identical to that of the leading vehicle. 
The powertrain operating points of the ego vehicles for the two methods are evaluated based on the battery electric vehicle powertrain efficiency map (Fig. \ref{fig:drive_powertrain_eff}). 
As shown in Fig. \ref{fig:Result_operation_point_compare}, the operating points of the CDFS are widely allocated with more points being found closer to the operational boundaries (red dash rectangle), which are less efficient regions. As for the REACC, the operating points are more concentrated and mostly located within the highly efficient zone. Therefore, the REACC method can drive the vehicle in a more ecological way compared to the CDFS, yielding 11.53\% ($N$=11 of REACC) of energy saving.

\begin{figure}[t!]
\centering
\includegraphics[width=\columnwidth]{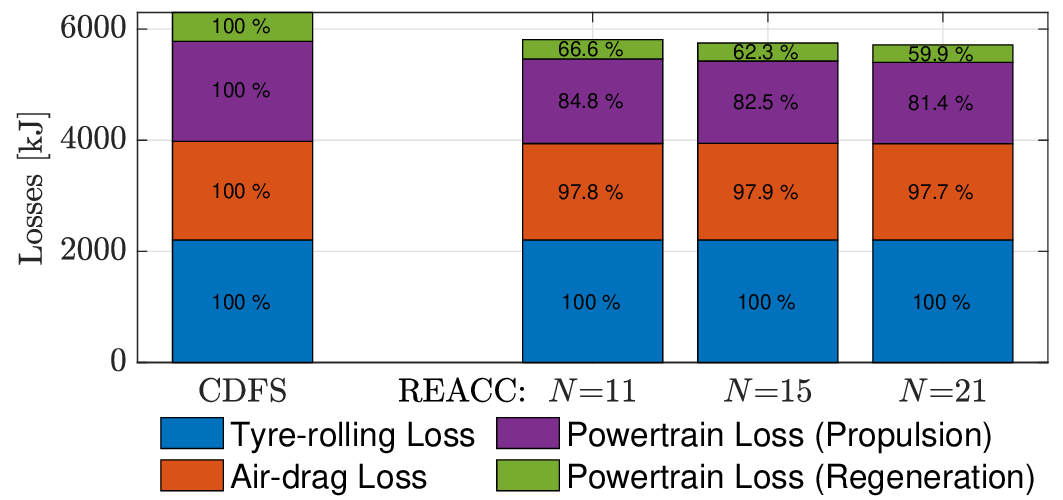}\\[-1ex]
\setlength{\belowcaptionskip}{-10pt}
\caption{Energy loss compositions breakdown by REACC ($N$=11,15,21) and CDFS schemes, for different prediction horizon lengths. 
Energy losses are categorised into four main types: tyre-rolling, air-drag, powertrain propulsion, and powertrain regeneration losses. Each type of loss is represented by a different colour and is labelled by a number that shows the relative quantity of the loss by REACC compared to the same type of loss by CDFS in terms of percentage. 
Although a portion of consumed energy can also be dissipated through mechanical braking, this type of loss is not included in the figure since its amount compared with other loss types is negligible. Specifically, mechanical braking amounts to $0.369\%$ of total losses for the CDFS scheme, while the REACC schemes can fully avoid mechanical braking losses.}
\label{fig:Result_losses_compare}
\end{figure}

Furthermore, the composition of various energy losses is presented in Fig. \ref{fig:Result_losses_compare}. 
Although the two schemes have similar amounts of air-drag and tyre-rolling resistance losses (illustrated by heights of the orange and blue bars, respectively), the proposed REACC can reduce up to $40.1\%$ and $18.6\%$ the energy losses caused by powertrain regeneration and propulsion, respectively. Besides, the mechanical braking is completely avoided when REACC is deployed, while it contributes to approximately $0.369\%$ of the losses in the case of the CDFS (not shown in Fig. \ref{fig:Result_losses_compare}), which reinforces the validity of the choice of \eqref{eq:space_multi_objective_convex_stage} as the OCP cost function, in which $F_t$ is replaced by $F_w$. As a result, the total powertrain loss (propulsion and regeneration) of REACC can be reduced by up to 23.43\% as compared to the CDFS scheme. Note that the change of kinetic energy corresponding to the difference between the initial and the final velocities of the ego vehicle, as well as the potential energy change because of the height differences between the initial and final positions of the ego vehicle are identical, respectively, in simulations with both the CDFS and REACC schemes due to the terminal speed condition imposed to the REACC scheme (similarly to Section \ref{sec:comparison_time_space}) and the equal travelled distance of the two schemes. Therefore, these quantities are not included in the energy loss comparison between the two schemes.
These findings further verify the finding in Fig.~\ref{fig:Result_operation_point_compare} of the capability of the REACC to achieve a more ecological driving behaviour. 
\subsection{Verification of the approach in real-time implementation}
\begin{figure}[t!]
\centering
\includegraphics[width=\columnwidth]{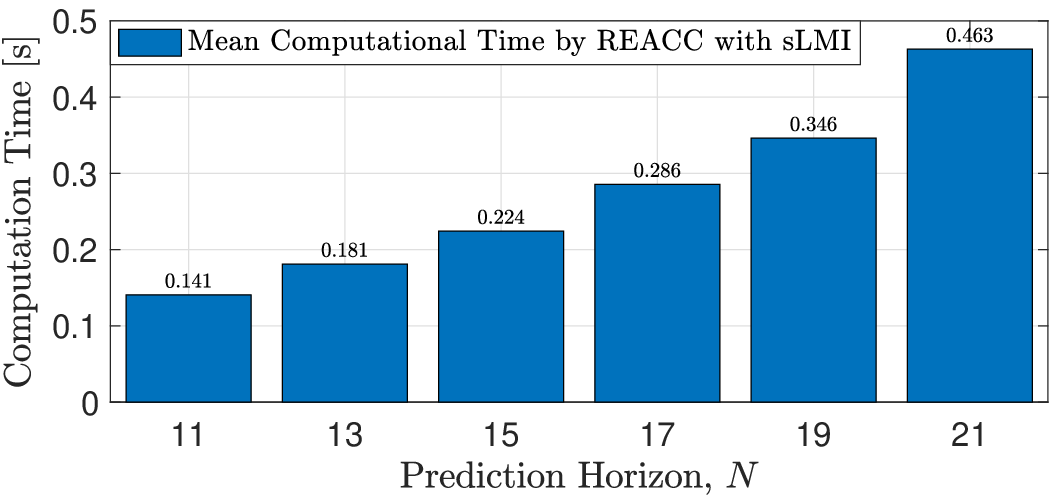}\\[-1ex]
\caption{Mean computation time per iteration for different prediction horizon lengths by REACC with the sLMI scheme.}
\setlength{\belowcaptionskip}{-10pt}
\label{fig:Result_time_compare}
\end{figure}

The computational burden of the proposed REACC is evaluated in this part. The test results are presented in Fig. \ref{fig:Result_time_compare}. The average computational time required for each iteration can be reduced to $0.141$~s when $N$=11. When the prediction horizon is enlarged, the computational cost monotonically increases and reaches $0.463$~s when $N$=21. Given the fixed sampling distance interval, $\delta s$=3~m, and the average velocity ($15.22$~m/s) of this experimental drive cycle, which would to an average sampling time interval of $0.21$~s, the result shows a possibility of implementing the REACC scheme in real-time. One of the key contributors to saving computation time is the sLMI, which is able to compact multiple LMIs into a single LMI. 

\section{Conclusion and Future Work}
\label{sec:conclusions}
In this work, the REACC problem is addressed by proposing a robust convex energy-optimal adaptive cruise control strategy. To deal with the disturbances dominated by modelling mismatches, an RMPC controller with SDPR and sLMI constraint formulation is designed to enhance the robustness of the system as well as address the computational issue. By applying conservative relaxation and convexification on the battery electric vehicle powertrain and the system dynamics, the overall REACC problem is formulated as a convex optimisation problem in the space-domain. 
The performance of the proposed REACC is evaluated on a real-world London UK driving cycle against three benchmarks under the same initial conditions and simulation disturbances.  
The robustness of the REACC is verified through the comparison with the infeasible solutions yielded by a nominal-based MPC.
As compared to a benchmark RMPC in the time-domain, the REACC method in space-domain is able to achieve a higher energy efficiency thanks to a precisely fitted battery powertrain model, as well as an improved driving comfort with less jerky manoeuvres, which further contributes to energy-saving.
Moreover, the REACC scheme is compared with a CDFS scheme by investigating energy economy, powertrain operating points and energy loss compositions, which further illustrates the contribution towards energy saving by the REACC strategy.
By further investigating computational time, it is verified that the REACC can be potentially implemented in real-time with enhanced energy economy, driving comfort and robustly satisfied safety constraints.

Future work could involve a further investigation of the performance and computational benefits of the proposed space-domain scheme in further realistic and more demanding driving missions as well as with consideration of practical powertrain characteristics, such as thermal effects. 

\begin{appendices}
\section{}
\label{app:sLMI Proof}
Recall the LMI in \eqref{eq:scr_L_substituted2}
\begin{equation}
    \begin{aligned}
       \!\!\begin{bmatrix}
           2\mu \!\! & \!\!\left(\mathbf{\overline{f}}^*\!\!-\!\mathbf{I}^*\!\!\left(\!\!\widetilde{C}_{\!f}\!x(0)\!+\!\widetilde{D}_{\!fu}\!\mathbf{u}\!+\!\widetilde{D}_{\!fc}\!\mathbf{C}\!+\!\widetilde{D}_{\!fd}\mathbf{d}\!\! \right)\!\!-\!\left(M\!e\!+\!e\mu \right)\!\!\right)^{\!\!\top}\\
           * & M+M^\top
       \end{bmatrix} \!\!\!\succeq \!0.   \nonumber
    \end{aligned}
\end{equation}
Applying Schur complement to transform the left-hand-side of $\eqref{eq:scr_L_substituted2}$ into a scalar
\begin{flalign*}
        =&2\mu-\\
        &\left(\mathbf{\overline{f}}^*\!-\!\mathbf{I}^*\!\left(\!\widetilde{C}_{\!f}x(0)\!+\!\widetilde{D}_{\!fu}\mathbf{u}\!+\!\widetilde{D}_{\!fc}\mathbf{C}\!+\!\widetilde{D}_{\!fd}\mathbf{d}\!\right)\!-\!\left(M\!e\!+\!e\mu \right)\!\right)^{\!\top}\!\cdot\\
        &\left(\!M\!+\!M\!^{\top}\!\right)^{\!-1}\cdot\\
        &\left(\mathbf{\overline{f}}^*\!-\!\mathbf{I}^*\!\left(\!\widetilde{C}_{\!f}x(0)\!+\!\widetilde{D}_{\!fu}\mathbf{u}\!+\!\widetilde{D}_{\!fc}\mathbf{C}\!+\!\widetilde{D}_{\!fd}\mathbf{d}\! \right)\!-\!\left(M\!e\!+\!e\mu \right)\!\right) \\\nonumber
        =&2\mu-\\
        &\left(\!\mathbf{\overline{f}}^{*\!\top}\!\!\!\!\!\!\!-\!\!
        x(0)^{\!\!\top}\!\widetilde{C}_{\!f}^{\top}\!\mathbf{I}^{*\!\top}\!\!\!\!\!-\!\mathbf{u}^{\!\top}\!\widetilde{D}_{\!f\!u}^{\top}\mathbf{I}^{*\!\top}\!\!\!-\!
        \mathbf{C}^{\top}\!\!\widetilde{D}_{\!f\!c}^{\top}\!\mathbf{I}^{*\!\top}\!\!\!\!\!-\!
        \mathbf{d}^{\!\top}\!\widetilde{D}_{\!f\!d}^{\top}\mathbf{I}^{*\!\top}
        \!\!\!-\!\left(\!M\!e\!+\!e\mu \right)^{\!\!\!\top}\!\!\right)\!\cdot\\
        &\left(M\!\!+\!\!M\!^{\top}\!\right)^{\!-1}\cdot\\
        &\left(\mathbf{\overline{f}}^*\!\!-\!\mathbf{I}^*\widetilde{C}_{\!f}\!x(0)\!-\!\mathbf{I}^*\widetilde{D}_{\!fu}\!\mathbf{u}\!-\!\mathbf{I}^*\widetilde{D}_{\!fc}\!\mathbf{C}\!-\!\mathbf{I}^*\widetilde{D}_{\!fd}\mathbf{d}\!-\!\left(M\!e\!+\!e\mu \right)\!\!\right)\!. \nonumber
\end{flalign*}
Since $\mathbf{\overline{f}}^{*}$, $\mathbf{I}^{*}\widetilde{C}_{f}x(0)$, and $\mathbf{I}^{*}\widetilde{D}_{fc}\mathbf{C}$ are all constant terms, by defining a new $\mathbf{\overline{f}}'=\mathbf{\overline{f}}^{*}-\mathbf{I}^*\widetilde{C}_{f}x(0)-\mathbf{I}^*\widetilde{D}_{fc}\mathbf{C}$,  the scalar can be further simplified as in
\begin{flalign*}
        =&2\mu-\left(\mathbf{\overline{f}}'^{\top}\!\!-\!\mathbf{u}^{\!\top}\widetilde{D}_{\!fu}^{\top}\mathbf{I}^{*\top}\!\!\!-
        \mathbf{d}^{\!\top}\!\widetilde{D}_{\!fd}^{\top}\mathbf{I}^{*\top}
        \!\!\!-\!\left(\!M\!e\!+\!e\mu \right)^{\!\!\top}\!\!\right)\cdot\\
        &\left(M\!\!+\!\!M\!^{\top}\!\right)^{\!-1}
        \left(\mathbf{\overline{f}}'\!-\!\mathbf{I}^{*}\widetilde{D}_{\!fu}\!\mathbf{u}\!-\!\mathbf{I}^{*}\widetilde{D}_{\!fd}\mathbf{d}\!-\!\left(M\!e\!+\!e\mu \right)\!\!\right) \\
      =&2\mu-\mathbf{\overline{f}}'^{\top}\left(M\!\!+\!\!M\!^{\top}\!\right)^{\!-1}\mathbf{\overline{f}}'+\mathbf{\overline{f}}'^{\top}\left(M\!\!+\!\!M\!^{\top}\!\right)^{\!-1}\!\mathbf{I}^{*}\widetilde{D}_{\!fu}\!\mathbf{u}+\\        &\mathbf{\overline{f}}'^{\top}\left(M\!\!+\!\!M\!^{\top}\!\right)^{\!-1}\!\mathbf{I}^{*}\widetilde{D}_{\!fd}\mathbf{d}+\mathbf{\overline{f}}'^{\top}\left(M\!\!+\!\!M\!^{\top}\!\right)^{\!-1}\left(M\!e\!+\!e\mu \right)+\\
        &\mathbf{u}^{\!\top}\widetilde{D}_{\!fu}^{\top}\mathbf{I}^{*\top}\!\!\left(M\!\!+\!\!M\!^{\top}\!\right)^{\!-1}\mathbf{\overline{f}}'
        -\mathbf{u}^{\!\top}\widetilde{D}_{\!fu}^{\top}\mathbf{I}^{*\top}\!\!\left(M\!\!+\!\!M\!^{\top}\!\right)^{\!-1}\!\mathbf{I}^{*}\widetilde{D}_{\!fu}\!\mathbf{u}-\\
        &\mathbf{u}^{\!\top}\widetilde{D}_{\!fu}^{\top}\mathbf{I}^{*\top}\!\!\left(M\!\!+\!\!M\!^{\top}\!\right)^{\!-1}\!\mathbf{I}^{*}\widetilde{D}_{\!fd}\mathbf{d}-\\
        &\mathbf{u}^{\!\top}\widetilde{D}_{\!fu}^{\top}\mathbf{I}^{*\top}\!\!\left(M\!\!+\!\!M\!^{\top}\!\right)^{\!-1}\!\!\left(M\!e\!+\!e\mu \right)+\\
        &\mathbf{d}^{\!\top}\!\widetilde{D}_{\!fd}^{\top}\mathbf{I}^{*\top}\!\!\left(M\!\!+\!\!M\!^{\top}\!\right)^{\!-1}\mathbf{\overline{f}}'
        -\mathbf{d}^{\!\top}\!\widetilde{D}_{\!fd}^{\top}\mathbf{I}^{*\top}\!\!\left(M\!\!+\!\!M\!^{\top}\!\right)^{\!-1}\!\mathbf{I}^{*}\widetilde{D}_{\!fu}\!\mathbf{u}-\\
        &\mathbf{d}^{\!\top}\!\widetilde{D}_{\!fd}^{\top}\mathbf{I}^{*\top}\!\!\left(M\!\!+\!\!M\!^{\top}\!\right)^{\!-1}\!\mathbf{I}^{*}\widetilde{D}_{\!fd}\mathbf{d}-\\
        &\mathbf{d}^{\!\top}\!\widetilde{D}_{\!fd}^{\top}\mathbf{I}^{*\top}\left(M\!\!+\!\!M\!^{\top}\!\right)^{\!-1}\left(M\!e\!+\!e\mu \right)+\\
        &\left(\!M\!e\!+\!e\mu \right)^{\!\!\top}\!\!\left(M\!\!+\!\!M\!^{\top}\!\right)^{\!-1}\mathbf{\overline{f}}'
        \!-\left(\!M\!e\!+\!e\mu \right)^{\!\!\top}\!\!\left(M\!\!+\!\!M\!^{\top}\!\right)^{\!-1}\!\mathbf{I}^{*}\widetilde{D}_{\!fu}\!\mathbf{u}-\\
        &\left(\!M\!e\!+\!e\mu \right)^{\!\!\top}\!\!\left(M\!\!+\!\!M\!^{\top}\!\right)^{\!-1}\!\mathbf{I}^{*}\widetilde{D}_{\!fd}\mathbf{d}-\\
        &\left(\!M\!e\!+\!e\mu \right)^{\!\!\top}\!\!\left(M\!\!+\!\!M\!^{\top}\!\right)^{\!-1}\!\!\left(M\!e\!+\!e\mu \right). \nonumber
\end{flalign*}
After applying the SDPR procedure to decouple terms with disturbance $\mathbf{d}$ and introducing a new variable, $0\!\preceq \tilde{D} \in \mathbb{D}^{2N} $, the scalar is rewritten as
\begin{equation}
        \!\!\!\!\!\!\!\!(\mathbf{d}-\underline{\mathbf{d}})^{\top} \tilde{D}(\overline{\mathbf{d}}-\mathbf{d})\!+\!\begin{bmatrix}\mathbf{d}^{\top}&1\end{bmatrix}L_{sLMI}(\tilde{D}, \mu, M, \mathbf{u})\begin{bmatrix}\mathbf{d}\\1\end{bmatrix}, \nonumber 
\end{equation}
where $ L_{sLMI}(\tilde{D}, \mu, M, \mathbf{u})$ is defined as
\begin{equation} \label{eq:app_sLMI_define_final}
    \begin{aligned}
    L_{sLMI}&(\tilde{D}, \mu, M, \mathbf{u})=\\
    &\begin{bmatrix}
            \tilde{D}& -\tilde{D}\overline{\mathbf{d}}\\ * &2 \mu + \underline{\mathbf{d}}^{\top}\tilde{D}\overline{\mathbf{d}}
        \end{bmatrix}-\\
        &\begin{bmatrix}\widetilde{D}_{fd}^{\top}\mathbf{I}^{*\top}\\\left(- \mathbf{\overline{f}}'+\mathbf{I}^{*}\widetilde{D}_{fu}\mathbf{u}+\left(Me\!\!+\!\!e\mu \right)\right)^{\top}\end{bmatrix}
        \left(M\!\!+\!\!M\!^{\top}\!\right)^{\!-1} \cdot\\
        &\begin{bmatrix}
             \mathbf{I}^{'}\widetilde{D}_{fd} & - \mathbf{\overline{f}}'+\mathbf{I}^{*}\widetilde{D}_{fu}\mathbf{u}+\left(Me\!\!+\!\!e\mu \right)
        \end{bmatrix}. \nonumber 
    \end{aligned}
\end{equation}
Furthermore, the Schur complement is utilised here to eliminate nonlinear terms included in the matrix. Moreover, substituting back $\mathbf{\overline{f}}'=\mathbf{\overline{f}}^{*}-\mathbf{I}^*\widetilde{C}_{f}x(0)-\mathbf{I}^*\widetilde{D}_{fc}\mathbf{C}$, as written in \eqref{eq:L_afterSDPR_linear3}, the matrix $L_{sLMI}(\tilde{D}, \mu, M, \mathbf{u})$ is 
\begin{equation}
    \begin{aligned}
        &L_{sLMI}(\tilde{D}, \mu, M, \mathbf{u})=\\
        &\!\!\!\!\!\begin{bmatrix}
            \tilde{D}\!\!& -\tilde{D}\overline{\mathbf{d}} & \widetilde{D}_{fd}^{\top}\mathbf{I}^{*\top}\\
            * \!\!&2 \mu \!\!+\!\! \underline{\mathbf{d}}^{\top}\!\!\tilde{D}\overline{\mathbf{d}} \!\!&\!\! \left(\!\!- \mathbf{\overline{f}}^*\!\!\!+\!\!\mathbf{I}^*\!\!\left(\widetilde{C}_{f}x(0)\!\!+\!\!\widetilde{D}_{fc}\mathbf{C}\!\!+\!\!\widetilde{D}_{fu}\mathbf{u}\!\right)\!\!+\!\!\left(Me\!\!+\!\!e\mu \right)\!\!\right)^{\!\!\!\top}\\
            *\!\!&*&  \left(M\!\!+\!\!M\!^{\top}\!\right)
        \end{bmatrix}\!. \nonumber
    \end{aligned}
\end{equation}
Hence, \eqref{eq:scr_L_substituted2} is held, if the LMI $L_{sLMI}(\tilde{D}, \mu, M, \mathbf{u}) \succeq 0$ is true.
\end{appendices}

\bibliographystyle{IEEEtran}
\bibliography{reference}

\end{document}